%% file: 4Q-phys-rev-v2.tex
\documentclass[aps,prd,twocolumn,groupedaddress,nofootinbib]{revtex4-1}
\usepackage{graphicx}
\usepackage{here}
\usepackage[dvips]{epsfig}
\def\beq{\begin{equation}}
\def\eeq{\end{equation}}
\def\bea{\begin{eqnarray}}
\def\eea{\end{eqnarray}}
\def\bq{\begin{quote}}
\def\eq{\end{quote}}

\def\nnb{\nonumber}
\def\ga{\left(}
\def\dr{\right)}
\def\aga{\left\{}
\def\adr{\right\}}

\def\rar{\rightarrow}

\def\nnb{\nonumber}

\def\la{\langle}
\def\ra{\rangle}
\def\nin{\noindent}
\def\ba{\vspace*{-0.2cm}\begin{array}}
\def\ea{\end{array}\vspace*{-0.2cm}}

\def\b{$\bullet~$}

\def\als{\alpha_s}

\def\gg2{\la\alpha_s G^2 \ra}
\def\gg3{g^3f_{abc}\la G^aG^bG^c \ra}
\def\ggg4{\la\als^2G^4\ra}

\def\GGG{\la G^3 \ra}
\def\gGGG{\la g^3G^3\ra}

\usepackage{amsmath}
\usepackage{slashed}
\usepackage{color}

\begin{document}

\title{Doubly-hidden  scalar  heavy molecules and  tetraquarks states from QCD at NLO}
\author{R.M. Albuquerque}
\affiliation{Faculty of Technology, Rio de Janeiro State University (FAT,UERJ), Brazil}
\email[Email address:~] {raphael.albuquerque@uerj.br}
\author{S. Narison
}
\affiliation{Laboratoire
Univers et Particules de Montpellier (LUPM), CNRS-IN2P3, \\
Case 070, Place Eug\`ene
Bataillon, 34095 - Montpellier, France\\
and\\
Institute of High-Energy Physics of Madagascar (iHEPMAD)\\
University of Ankatso,
Antananarivo 101, Madagascar}
\email[Email address:~] {snarison@yahoo.fr}
\author{A. Rabemananjara}
\email[Email address:~] {achrisrab@gmail.com}
\author{D. Rabetiarivony}
\email[Email address:~] {rd.bidds@gmail.com}
\author{G. Randriamanatrika}
\email[Email address:~] {artesgaetan@gmail.com}

\affiliation{Institute of High-Energy Physics of Madagascar (iHEPMAD)\\
University of Ankatso,
Antananarivo 101, Madagascar}

\date{\today}
\begin{abstract}
\noindent
 Alerted by the recent LHCb discovery of exotic hadrons in the range (6.2    6.9) GeV, we present new results for the doubly-hidden scalar heavy  $(\bar QQ) (Q\bar Q)$ charm and beauty molecules using the inverse Laplace transform sum rule (LSR)  within stability criteria and including the Next-to-Leading Order (NLO) factorized perturbative and $\la G^3\ra$ gluon condensate corrections. We also critically revisit and improve existing Lowest Order (LO) QCD spectral sum rules (QSSR) estimates of the $({ \bar Q \bar Q})(QQ)$ tetraquarks analogous states. 
In the example of the anti-scalar-scalar molecule, 
we separate explicitly the contributions of the factorized and non-factorized contributions to LO of perturbative QCD and to the $\la\alpha_sG^2\ra$ gluon condensate contributions in order to disprove some criticisms on the (mis)uses of the sum rules for four-quark currents. 
We also re-emphasize the importance to include PT radiative corrections for heavy quark sum rules in order to justify the (ad hoc) definition and value of the heavy quark mass used frequently  at LO in the literature. 
Our LSR results for tetraquark masses summarized in Table\,\ref{tab:res} are compared with the ones from ratio of moments (MOM) at NLO and results from LSR and ratios of MOM at LO (Table\,\ref{tab:comparison}). The LHCb broad structure around (6.2  --  6.7) GeV can be described by the $\overline{\eta}_{c}{\eta}_{c}$, $\overline{J/\psi}{J/\psi}$ and $\overline{\chi}_{c1}{\chi}_{c1}$ molecules or/and their analogue tetraquark scalar-scalar, axial-axial and vector-vector lowest mass ground states. The peak at (6.8    -- 6.9) GeV can be likely due to a $\overline{\chi}_{c0}{\chi}_{c0}$ molecule or/and a pseudoscalar-pseudoscalar tetraquark state. Similar analysis is done for  the scalar beauty states whose masses are found to be above the $\overline\eta_b\eta_b$ and $\overline\Upsilon(1S)\Upsilon(1S)$ thresholds. 
\end{abstract}
\maketitle
 \section{Introduction}
QCD spectral sum rules (QSSR)\`a la SVZ\,\cite{SVZa,SVZb,ZAKA} have been applied 
since 41 years\,\footnote{For revieews, see e.g\,\cite{SNB1,SNB2,SNB3,SNB4,SNB5,SNREV15,SNREV10,IOFFEb,RRY,DERAF,BERTa,YNDB,PASC,DOSCH}.}  to study successfully the hadron properties (masses, couplings and widths) and to extract some fundamental QCD parameters ($\alpha_s$, quark masses, quark and gluon condensates,...). In previous series of papers\,\cite{HEP18,SU3,QCD16,MOLE16,X5568,MOLE12}, we have used the inverse Laplace transform (LSR)\,\cite{BELLa,BELLb,BECCHI,SNR}  of  QSSR to predict the couplings and masses of different heavy-light molecules and tetraquarks states by including next-to-next nonleading order (N2LO) factorized perturbative (PT) corrections where we have emphasized the importance of these corrections for giving a meaning of the input heavy quark mass which plays an important role in the analysis though these corrections are small in the $\overline{MS}$-scheme. However, this feature (a posteriori) can justify the uses of the  $\overline{MS}$ running masses at LO in some channels\,\cite{MOLEREV} if the $\alpha_s^n$-corrections are small, especially in the ratios of moments used to extract the hadron masses where these corrections tend to compensate\cite{SNB1,SNB2}.

In this paper, we pursue the analysis for the fully\,/\,doubly-hidden heavy quarks $(\bar QQ) (Q\bar Q)$ molecules and $( \bar Q \bar Q)(QQ)$ tetraquarks states, where the effect of the quark mass value and its definition are (a priori) important as we have four heavy quarks which bound these states.  

We separate explicitly the factorized and non-factorized contributions to the four-quark correlators at LO of PT QCD and for the lowest dimension gluon condensate $\la \alpha_s G^2\ra$ contributions. We add the contribution of the NLO perturbative corrections  from the factorized part of the diagrams which as we shall see is a good approximation. We also include the triple gluon condensate $\la G^3\ra$
contributions in the Operator Product Expansion (OPE). 

We use these QCD results using the LSR sum rules within different stability criteria used successfully in some other channels to extract the masses and couplings of the previous molecules and tetraquarks states assumed to be resonances. 

Our results as improved estimates of the LO ones given in the QSSR literature.

We expect that these will be an useful guide for further experimental searches of these exotic states and for identifying the  different new states found by LHCb\,\cite{LHCb1,LHCb2}. 


\section{The Inverse Laplace sum rules}
\subsection{The QCD molecule local interpolating currents}
    We shall be concerned with the following QCD local interpolating currents of dimension-six:
\beq
\hspace*{-0.6cm}\la 0|{\cal O}^H_{\cal M}(x)\vert {\cal M}\ra= f_{\cal M}^HM_{\cal M}^4:~ {\cal O}^H_{\cal M}(x)\equiv (J_{\cal M}^{H}\bar J_{\cal M}^H)(x)
\eeq
where  $f^H_{\cal M}$ is the meson decay constant; $ J^{H}_{\cal M}(x)$ is the lowest dimension bilinear quark currents and $H\equiv S,P,V,A$. 

   For the scalar ($0^{++}$) molecule states, these currents are\,:
\beq
J^{[S,P,V,A]}_{\cal M}=\bar Q[1,\gamma_5,\gamma_\mu, \gamma_5\gamma_\mu]Q~.
\label{eq:mole}
\eeq
Interpolating currents constructed from bilinear (pseudo)scalar currents are not renormalization group invariants such that the corresponding  decay constants possess anomalous dimension:
\beq
f^{(S,P)}_{\cal M}(\mu)=\hat f^{(S,P)}_{\cal M} \ga -\beta_1a_s\dr^{4/\beta_1}(1-k_fa_s) ,
\eeq
where : $\hat f^{(S,P)}_{\cal M}$ is the renormalization group invariant coupling and $-\beta_1=(1/2)(11-2n_f/3)$ is the first coefficient of the QCD $\beta$-function for $n_f$ flavours. $a_s\equiv (\alpha_s/\pi)$ is the QCD coupling. $k_f=2.028 (2.352)$ for $n_f=4(5)$ flavours. 
\subsection{Form of the sum rules}
We shall work with the  Finite Energy version of the QCD Inverse Laplace sum rules (LSR) and their ratios :
\bea
 {\cal L}^c_n(\tau,\mu)&=&\int_{16m_Q^2}^{t_c}\hspace*{-0.5cm}dt~t^n~e^{-t\tau}\frac{1}{\pi} \mbox{Im}~\Pi^H_{\cal M}(t,\mu)~,\nnb\\
 {\cal R}^c_n(\tau)&=&\frac{{\cal L}^c_{n+1}} {{\cal L}^c_n},
\label{eq:lsr}
\eea
 where $m_Q$ is the heavy quark mass, $\tau$ is the LSR variable, $n=0,1$ is the degree of moments, $t_c$ is  the threshold of the ``QCD continuum" which parametrizes, from the discontinuity of the Feynman diagrams, the spectral function  ${\rm Im}\,\Pi^H_{\cal M}(t,m_Q^2,\mu^2)$   where  $\Pi^H_{\cal M}(t,m_Q^2,\mu^2)$ is the  scalar correlator defined as :
 \beq
\hspace*{-0.6cm} \Pi^H_{\cal M}(q^2)=\int \hspace*{-0.15cm}d^4x ~e^{-iqx}\la 0\vert {\cal T} {\cal O}^H_{\cal M}(x)\ga {\cal O}^H_{\cal M}(0)\dr^\dagger \vert 0\ra~.
 \label{eq:2-pseudo}
 \eeq
 \\
\section{The QCD two-point function}
 \subsection{The LO $\oplus$ $\la G^2\ra$ contributions}
\begin{figure}[hbt]
\vspace*{-0.75cm}
\begin{center}
\hspace*{0.5cm}\includegraphics[width=8cm]{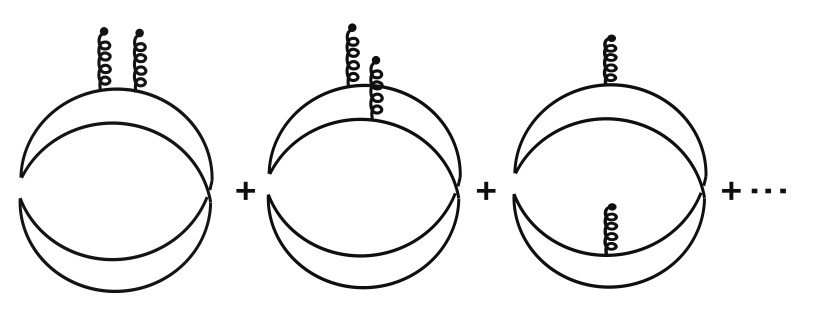}
\vspace*{-0.25cm}
\caption{\footnotesize  Non-factorised $\la G^2\ra$ contribution to the spectral function.} 
\label{fig:g2nf}
\end{center}
\end{figure}

Using the SVZ\,\cite{SVZa,SVZb} Operator Product Expansion (OPE), we give below the QCD expression of the two-point correlators associated to the $\overline\chi_0\chi_0$ molecule to LO of PT QCD and up to dimension-four condensates can be extracted from the Feynman diagrams in Figs.\,\ref{fig:pert} to\,\ref{fig:g2nf} : 
\begin{figure}[hbt]
\vspace*{-0.75cm}
\begin{center}
\includegraphics[width=2.5cm]{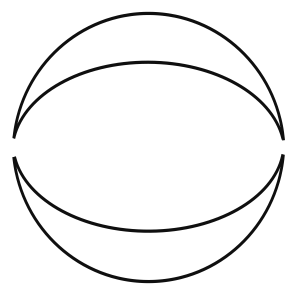}\hspace*{1cm}
\includegraphics[width=2.5cm]{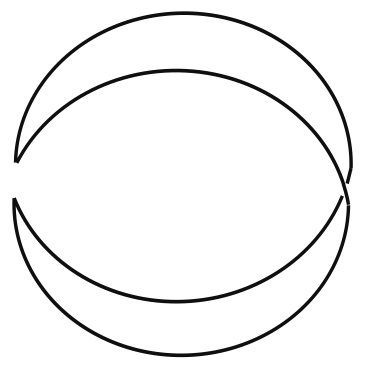}
\centerline {\hspace*{0.cm} \bf (a) \hspace*{3cm}\bf (b)}
\vspace*{-0.25cm}
\caption{\footnotesize  LO PT contribution to the spectral function : (a) factorised diagram ; (b) non-factorised diagram.} 
\label{fig:pert}
\end{center}
\end{figure}
\begin{figure}[hbt]
\vspace*{-0.5cm}
\begin{center}
\hspace*{0.5cm}\includegraphics[width=8cm]{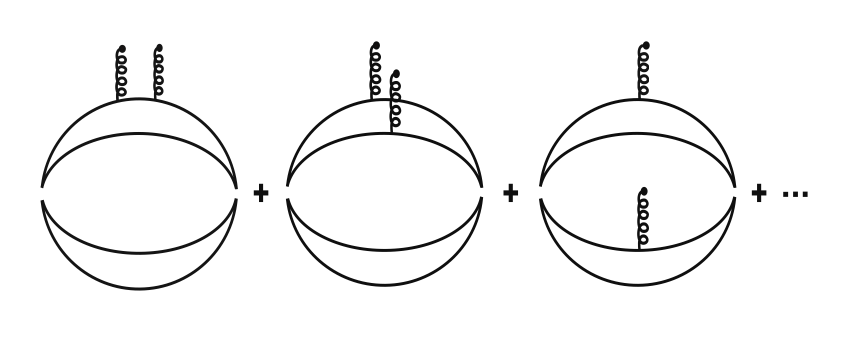}
\vspace*{-0.25cm}
\caption{\footnotesize  Factorised $\la G^2\ra$ contribution to the spectral function where the 3rd diagram gives a null contribution. } 
\label{fig:g2f}
\end{center}
\end{figure}
\begin{widetext}
\vspace*{-0.5cm}
\bea
\frac{1}{\pi}{\rm Im}\,\Pi^{S;\,LO}_{\overline\chi_{0q}\chi_{0q}}(t)&=&\frac{3}{2^{9}\pi ^6}\int_{x\,y\,z} \mathcal{F}_2(M^2,t) \left[6 m^4+4\,x\,y\,m^2 (2M^2-5t)-3\,x\,y\,z\,(x+y+z-1)(M^4-6 M^2 t+7t^2)\right]\nnb\\
&&-\frac{\bf\epsilon}{2^{11}\pi ^6}\int_{x\,y\,z} \mathcal{F}_2(M^2,t) \left[6 m^4\hspace*{-0.1cm}+4\,x\,y\,m^2 (2M^2-5t)-3\,x\,y\,z\,(x+y+z-1)(M^4-6 M^2 t+7t^2)\right]\nnb\\
\frac{1}{\pi}{\rm Im}\,\Pi^{S;\,G^2}_{\overline\chi_{0q}\chi_{0q}}(t)&=&\frac{\la\alpha_s  G^2\ra}{2^7 \pi^5} \int_{x\,y\,z} \frac{1}{x^3 y} \Big\{3 x^2 \big[2 m^4+m^2 \big(2 M^2-3 t\big) \big(x\,(y-2 z)-2 z\,(y+z-1)\big)-x\,y\,z\, \big(x+y+z-1\big),\nnb\\
&&\times\big(3 M^4-12 M^2 t+10 t^2\big)\big]
+2 m^2 y\,\big[3 x\,z\,\big(x+y+z-1\big)\big(t\,(4 y+3)-2 M^2 (y+1)\big)+ m^2 \big(x\,(2 y - 2 z + 3),\nnb\\
&&-2 z\,(y+z-1)\big)+
\big(m^2-t\,x\,y\big) \big(m^2+t\,z\,(x+y+z-1)\big)\,\delta\,\big(t-M^2\big)\big]\Big\},\nnb \\
&&+\frac{{\bf\epsilon}\, \la \alpha_s G^2\ra}{3\times 2^{9} \pi^5} \int_{x\,y\,z}\frac{1}{x^3 y} \Big\{3 x^2 \big[-6 m^4 - m^2 \big(2 M^2-3 t\big) \big(x\,y-4 z\,(y+z-1)\big)+3x\,y\,z\, \big(x+y+z-1\big)\nnb\\
&&\times\big(3 M^4-12 M^2 t+10 t^2\big)\big]+2 m^2 y\,\big[3 x\,z\,\big(x+y+z-1\big)\big(2 M^2 (y+1)-t\,(4 y+3)\big)+ m^2 \big(2 z\,(y+z-1)\nnb\\
&&-x\,(2 y - 2 z + 3)\big)-\big(m^2-t\,x\,y\big) \big(m^2+t\,z\,(x+y+z-1)\big)\,\delta\,\big(t-M^2\big)\big]\Big\}~,
\vspace*{-0.25cm}
\label{eq:scalar}
\eea

where  $m \equiv m_Q$~\rm{is the heavy quark mass}, $\epsilon$=0 corresponds to the factorized contribution and $\epsilon$ =1 to the sum of factorized $\oplus$ non-factorized ones. The other parameters are :
 \bea
 x_{_{\substack{max \\ min}}}&=&\frac{1}{2} \aga\left(1-\frac{8 m^2}{t}\right)\pm \sqrt{\left(1-\frac{8m^2}{t}\right)^2-\frac{4 m^2}{t}}\adr,\nnb\\
 y_{_{\substack{max \\ min}}}&=&\frac{1}{2} \aga 1\pm\left[\sqrt{\frac{\left(m^2+t (x-1) x\right) \left(m^2 (8 x+1)+t (x-1) x\right)}{\left(m^2-t x\right)^2}}+x \left(\frac{3 t x}{m^2-t x}+2\right)\right]\adr,\nnb\\
 z_{_{\substack{max \\ min}}}&=&\frac{1}{2} \aga (1-x-y)\pm\sqrt{\frac{(x+y-1) \left(m^2 \left(-x^2+2 x y+x-y^2+y\right)+t x y (x+y-1)\right)}{t x y-m^2 (x+y)}}\adr,\nnb\\
\frac{M^2}{m^2}&=&\frac{ 1}{x}+\frac{1}{y}+\frac{1}{z}+\frac{1}{1-x-y-z},~~
 \mathcal{F}_n(M^2,t) = (M^2 - t)^n,~~
 Q^2=-q^2 ,
  \int_{x\,y\,z}  \hspace*{-0.4cm} \equiv \int^{x_{max}}_{x_{min}} \hspace*{-0.6cm} {\rm d}x\, \int^{y_{max}}_{y_{min}} \hspace*{-0.6cm} {\rm d}x\, \int^{z_{max}}_{z_{min}}\hspace*{-0.6cm}\,{\rm d}z. 
  \label{eq:var}
 \eea
  \end{widetext}

   The contribution of the $\la g^3 G^3\ra$ condensate is quite lengthy and is given in Appendix A.

   We have cross-checked that, using our calculation method, we recover the results for charmonium where the heavy quark condensate contribution is already included into the gluon condensate one through the relation\,\cite{SVZa,SVZb,GENERALIS1,GENERALIS2,BAGAN1,BAGAN2}:
\beq
m_Q\la\bar QQ\ra= -\frac{1}{12\pi}\la\alpha_s G^2\ra-\frac{1}{1440\pi^3}\frac{\la g^3G^3\ra}{m^2_Q}+\cdots
\eeq

   The LO $\oplus$ $\la G^2\ra$ expressions of the other molecules spectral functions are given in  Appendix B. The one of the $\la G^3\ra$ condensates which are lengthy are not quoted. 

   We note that the inclusion of higher dimension condensate contributions ($d\geq 8$) , as abusively done in the current literature, does not help, except in some few cases, because the OPE is often convergent at the optimization scale while the size of higher dimension condensates are not under control due to the violation of factorization for the four-quark\,\cite{SNTAU,JAMI2,LNT,LAUNERb} and to the inaccuracy of the dilute gas instanton estimate of higher dimensions gluon\,\cite{SNH10,SNH11,SNH12} condensates.

\subsection{NLO PT corrections to the Spectral functions}

\begin{figure}[hbt]
\vspace*{-0.5cm}
\begin{center}
\hspace*{0.5cm}\includegraphics[width=6cm]{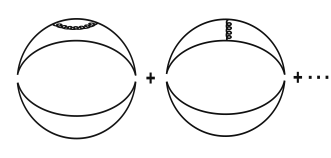}
\vspace*{-0.25cm}
\caption{\footnotesize  NLO factorised PT contribution to the spectral function. } 
\label{fig:as}
\end{center}
\end{figure}

We extract the next-to-leading (NLO) perturbative (PT) corrections by considering that the molecule /tetraquark two-point spectral function is the convolution of the two ones built from 
two quark bilinear currents (factorization) as illustrated in Fig.\,\ref{fig:as}. This is a good approximation because we have seen for the LO that the non-factorized part of the QCD diagrams gives a small contribution and behaves like 1/$N_c$ where $N_c$ is the number of colours. 
\bea\label{eq:qqcurrent}
J^{P,S}(x)&\equiv&\bar Q [i\gamma_5,1] Q ~~~\rar ~~~ \frac{1}{\pi}{\rm Im}\,\psi^{P,S}(t) \nnb\\
J^{V,A}(x)&\equiv&\bar Q [\gamma_\mu,\gamma_\mu\gamma_5] Q ~~~ \rar ~~~ \frac{1}{\pi}{\rm Im}\,\psi^{V,A}(t)
\eea
In this way, we obtain the convolution integral\,\cite{PICH,SNPIVO}:
\bea
\hspace*{-0.1cm}\frac{1}{ \pi}{\rm Im}\, \Pi^H_{\cal M}(t)&=& \theta (t-16m_Q^2)\ga  \frac{1}{4\pi}\dr^2\hspace*{-0.15cm} t^2 \hspace*{-0.15cm}\int_{4m_Q^2}^{(\sqrt{t}-2m_Q)^2}\hspace*{-0.8cm}dt_1\times\nnb\\
&&\int_{4m_Q^2}^{(\sqrt{t}-\sqrt{t_1})^2}  \hspace*{-0.8cm}dt_2~\lambda^{1/2}{\cal K}^H,
\eea
where : 
\bea
{\cal K}^{S,P}\hspace*{-0.1cm}&\equiv&\hspace*{-0.1cm}\ga \frac{t_1}{ t}+ \frac{t_2}{ t}-1\dr^2
\times ~ \frac{1}{\pi}{\rm Im} \,\psi^{S,P}(t_1) \frac{1}{\pi} {\rm Im}\, \psi^{S,P}(t_2),\nnb\\
\hspace*{-0.3cm}{\cal K}^{V,A}\hspace*{-0.1cm}&\equiv&\hspace*{-0.1cm}\Bigg{[}\ga \frac{t_1}{ t}+ \frac{t_2}{t}-1\dr^2 \hspace*{-0.15cm}+8\frac{t_1t_2}{ t^2}\Bigg{]}
\times \frac{1}{\pi}{\rm Im} \,\psi^{V,A}(t_1) \times\nnb\\
&& \frac{1}{\pi} {\rm Im}\, \psi^{V,A}(t_2),
\eea
with the phase space factor:
\beq
\lambda=\ga 1-\frac{\ga \sqrt{t_1}- \sqrt{t_2}\dr^2}{ t}\dr \ga 1-\frac{\ga \sqrt{t_1}+ \sqrt{t_2}\dr^2}{ t}\dr~,
\eeq
and $m_Q$ is the on-shell / pole perturbative heavy quark mass.

   The NLO perturbative expressions of the bilinear equal masses pseudoscalar spectral functions are known in the literature \cite{BROAD,RRY,SNB1,SNB2}. 

   We estimate the N2LO contributions assuming a geometric growth of the numerical coefficients\,\cite{SZ}. We consider this contribution as an estimate of the error due to the truncation of the PT series.

\subsection{From the On-shell to the $\overline{MS}$-scheme}
We transform the pole masses $m_Q$ to the running masses $\overline m_Q(\mu)$ using the known relation  in the
$\overline{MS}$-scheme to order $\alpha_s^2$ \cite{TAR,COQUEa,COQUEb,SNPOLEa,SNPOLEb,BROAD2a,BROAD2b,CHET2a,CHET2b}:
\bea
m_Q &=& \overline{m}_Q(\mu)\Big{[}
1+\frac{4}{3} a_s+ (16.2163 -1.0414 n_l)a_s^2\nnb\\
&&+\ln{\frac{\mu^2}{ \overline{m}_Q^2}} \ga a_s+(8.8472 -0.3611 n_l) a_s^2\dr\nnb\\
&&+\ln^2{\frac{\mu^2}{ \overline{m}_Q^2}} \ga 1.7917 -0.0833 n_l\dr a_s^2...\Big{]},
\label{eq:pole}
\eea
for $n_l=3: u,d,s$ light flavours. In the following, we shall use $n_f$=4 or 5 total number of flavours for the numerical value of $\alpha_s$ respectively for the charm and bottom quarks. 
\section{QCD input parameters}
\nin
The QCD parameters which shall appear in the following analysis will be the QCD coupling $\alpha_s$ the charm and bottom quark masses $m_{c,b}$,
 the gluon condensates $ \la\alpha_sG^2\ra$.  Their values are given in Table\,\ref{tab:param}. 
{\scriptsize
\begin{table}[hbt]
\setlength{\tabcolsep}{0.05pc}
    {\small
  \begin{tabular}{llll}
&\\
\hline
\hline
Parameters&Values&Sources& Ref.    \\
\hline
$\alpha_s(M_Z)$& $0.1181(16)(3)$&$M_{\chi_{0c,b}-M_{\eta_{c,b}}}$&LSR \, \cite{SNparam}\\
$\overline{m}_c(\overline {m}_c)$&$1286(16)$ MeV&$B_c\oplus {J/\psi}$&Mom.\,\cite{SNbc20,SNmom18}\\
$\overline{m}_b(\overline {m}_b)$&$4202(8)$ MeV&$B_x\oplus{\Upsilon}$&Mom.\,\cite{SNbc20,SNmom18}\\
$\la\alpha_s G^2\ra\times 10^2$& $(6.35\pm 0.35)$ GeV$^2$&Hadrons&Average\,\cite{SNparam}\\
$\la g^3  G^3\ra/\la\alpha_s G^2\ra$& $(8.2\pm 1.0)$ GeV$^2$&$J/\psi$  family&QSSR \cite{SNH10,SNH11,SNH12}\\
\hline\hline
\end{tabular}
}
 \caption{QCD input parameters from recent QSSR analysis based on stability criteria.
 $\overline{m}_{c,b}(\overline {m}_{c,b})$ are the running $c,b$ quark masses evaluated at $\overline{m}_{c,b}$. }  
\label{tab:param}
\end{table}
} 
\vspace*{-0.5cm}
\subsection{QCD coupling $\alpha_s$}
We shall use from the $M_{\chi_{0c}}-M_{\eta_{c}}$ mass-splitting sum rule\,\cite{SNparam}: 
 \bea&&\hspace*{-1cm} 
 \alpha_s(2.85)=0.262(9) \rar\alpha_s(M_\tau)=0.318(15)\nnb\\
& \rar&\alpha_s(M_Z)=0.1183(19)(3)
\eea
which is more precise than the one from  $M_{\chi_{0b}}-M_{\eta_{b}}$\,\cite{SNparam} : 
\bea &&\hspace*{-1cm} 
 \alpha_s(9.50)=0.180(8) \rar\alpha_s(M_\tau)=0.312(27)\nnb\\
&&\rar\alpha_s(M_Z)=0.1175(32)(3).
 \eea
 These lead to the mean value quoted in Table\,\ref{tab:param}, which is
 in complete agreement with the world average\,\cite{PDG}:
\beq
\alpha_s(M_Z)=0.1181(11)~.
\eeq
\subsection{$c$ and $b$ quark masses}
For the $c$ and $b$ quarks, we shall use the recent determinations\,\cite{SNmom18,SNbc20} of  the running masses and the corresponding value of $\alpha_s$ evaluated at the scale $\mu$ obtained using the same sum rule approach from charmonium and bottomium systems. 
\subsection{Gluon  condensate $\la \alpha_s G^2\ra$}
We use the recent estimate obtained from a correlation with the values of the heavy quark masses and $\alpha_s$ which can be compared with the QSSR average from different channels\,\cite{SNparam}.  


\section{The spectral function\label{sec:spectral}}
   In the present case, where no complete data on the spectral function are available, we use the duality ansatz:
 \bea
\hspace*{-0.65cm} {\rm Im} \Pi^H_{\cal M}\hspace*{-0.1cm}&\simeq&\hspace*{-0.1cm} f_H^2 M_H^{4} \delta(t-M_H^2) +\Theta(t-t_c) ``{\rm  Continuum}",
 \eea
 for parametrizing the molecule spectral function. $M_H$ and $f_H$ are the lowest ground state mass and coupling analogue to $f_\pi$. The ``Continuum" or ``QCD continuum"  is the imaginary part of the QCD correlator from the  threshold $t_c$. Within a such parametrization, one  obtains: 
 \beq
  {\cal R}^{c}_n\equiv {\cal R}\simeq M_H^2~,
  \label{eq:mass}
  \eeq
 indicating that the ratio of moments appears to be a useful tool for extracting the mass of the hadron ground state\,\cite{SNB1,SNB2,SNB3,SNB4,SNREV15}. 
 
    This simple model has been tested in different channels where complete data are available (charmonium, bottomium and $e^+e^-\to I=1$ hadrons)\,\cite{SNB1,SNB2,BERTa}. It was shown that, within the model, the sum rule  reproduces well the one using the complete data, while
the masses of the lowest ground state mesons ($J/\psi,~\Upsilon$ and $\rho$) have been predicted with a good accuracy.  In the extreme case of the Goldstone pion, the sum rule using the spectral function parametrized by this simple model\,\cite{SNB1,SNB2} and the more complete one by ChPT\,\cite{BIJNENS} lead to  similar values of the sum of light quark masses $(m_u+m_d)$ indicating the efficiency of this simple parametrization. 

   An eventual violation of the quark-hadron duality (DV)\,\cite{SHIF,PERIS,BOITO} has been frequently tested  in the accurate determination of $\alpha_s(\tau)$ from hadronic $\tau$-decay data\,\cite{PERIS,SNTAU,PICHROD}, where its quantitative effect in the spectral function was found to be less than 1\%. Typically, the DV behaves as: 
\beq
\Delta{\rm Im} \Pi^H_{\cal M,T}(t)\sim t~{\rm e}^{-\kappa t} {\rm sin} (\alpha+\beta t) \theta (t-t_c)~,
\eeq 
where $\kappa,\alpha,\beta$ are model-dependent  fitted parameters but not based from first principles. Within this model, where the contribution is doubly exponential suppressed in the Laplace sum rule analysis, we expect that in the stability regions where the QCD continuum contribution to the sum rule is minimal and where the optimal results in this paper will be extracted, such duality violations can be safely neglected. 

   Therefore, we (a priori) expect that one can extract with a good accuracy the  masses and  decay constants of the  mesons within the approach. An eventual improvement of the results can be done after a more complete measurement of the corresponding spectral function which is not an easy experimental task. 
 
   In the following, in order to minimize the effects of unkown higher radial excitations smeared by the QCD continuum and some eventual quark-duality violations, we shall work with the lowest ratio of moments $ {\cal R}^{c}_0$ for extracting the meson masses and with the lowest moment $ {\cal L}^c_0 $ for estimating the decay constant $f_{H}$. Moment with negative $n$ will not be considered due to their sensitivity on the non-perturbative contributions at zero momentum. 
  
\section{Optimization Criteria}
   For extracting the optimal results from the analysis, we have used in previous works the optimization criteria (minimum sensitivity) of the observables versus the variation  of the external variables namely the $\tau$ sum rule parameter, the QCD continuum threshold $t_c$ and the subtraction point $\mu$. 

   Results based on these criteria have lead to successful predictions in the current literature\,\cite{SNB1,SNB2}. $\tau$-stability has been introduced and tested by Bell-Bertlmann using the toy model of harmonic oscillator\,\cite{BERTa} and applied successfully in the heavy\, \cite{BELLa,BELLb,BERTa,BERTb,BERTc,BERTd,NEUF,SHAW,SNcb3,SNHeavy,SNHeavy2,SNHQET13}  and light quarks systems\,\cite{SVZa,SVZb,SNB1,SNB2,SNB3,SNB4,SNREV15,SNL14}. 

   It has been extended later on to the $t_c$-stability\,\cite{SNB1,SNB2,SNB3,SNB4} and  to the $\mu$-stability criteria\,\cite{SNp13,SNHQET13,SNL14,SNp15,SNparam}. 

   Stability on the number $n$ of heavy quark moments have also been used\,\cite{SNH12,SNH11,SNH10,SNmom18}. 

   One should notice in the previous works that these criteria have lead to more solid theoretical basis and noticeable improvement of the sum rule results. The quoted errors in the results are conservative as the range covered by $t_c$ from the beginning of $\tau$-stability to the one of $t_c$-stability is quite large. However, such large errors induce less accurate predictions compared with some other approaches (potential models, lattice calculations) especially for the masses of the hadrons. 
This is due to the fact that, in most cases, there are no available data for the radial excitations which can be used to restrict the range of $t_c$-values.  However, the value of $t_c$ used in the ``QCD continuum" model does not necessarily coincide with the 1st radial excitation mass as the "QCD continuum" is expected to smear all higher states contributions to the spectral function. This feature has been explicitly verified by\,\cite{LAUNERb} in the $\rho$-meson channel.



\section{The scalar ${{\overline{\chi}_{c0}}\chi_{c0}}$  molecule\,\label{sec:chi0}}
Using the previous QCD expression given in Eq.\,\ref{eq:scalar} and adding the PT NLO contribution, we study the dependence of the coupling and mass on the LSR parameter $\tau$, the continuum threshold $t_c$ and the subtraction scale $\mu$. We shall also study the relative contribution of the continuum versus the ground state one. 
\subsection{$\tau$- and $t_c$-stabilities}
\begin{figure}[hbt]
\vspace*{-0.25cm}
\begin{center}
\includegraphics[width=8cm]{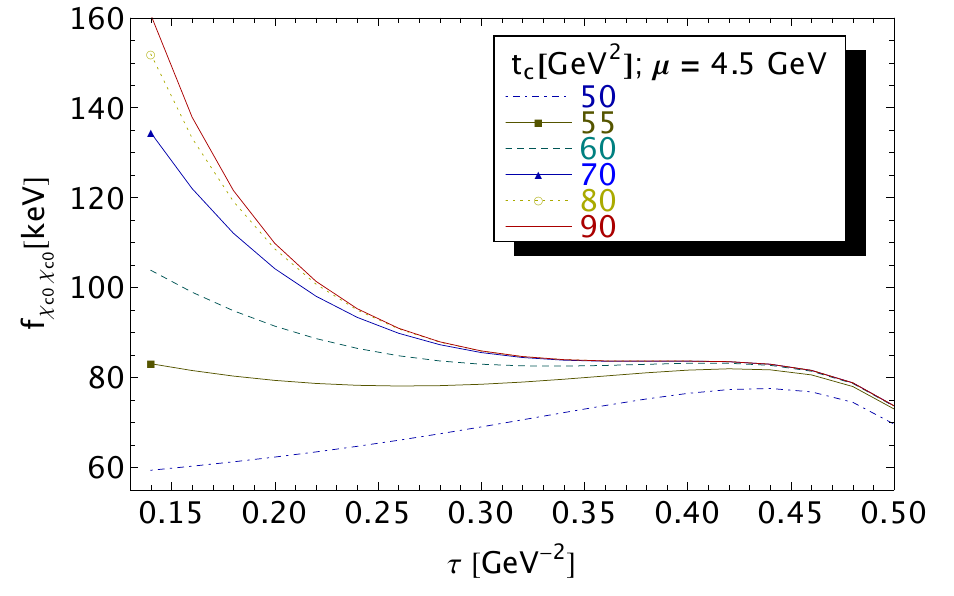}
\includegraphics[width=8cm]{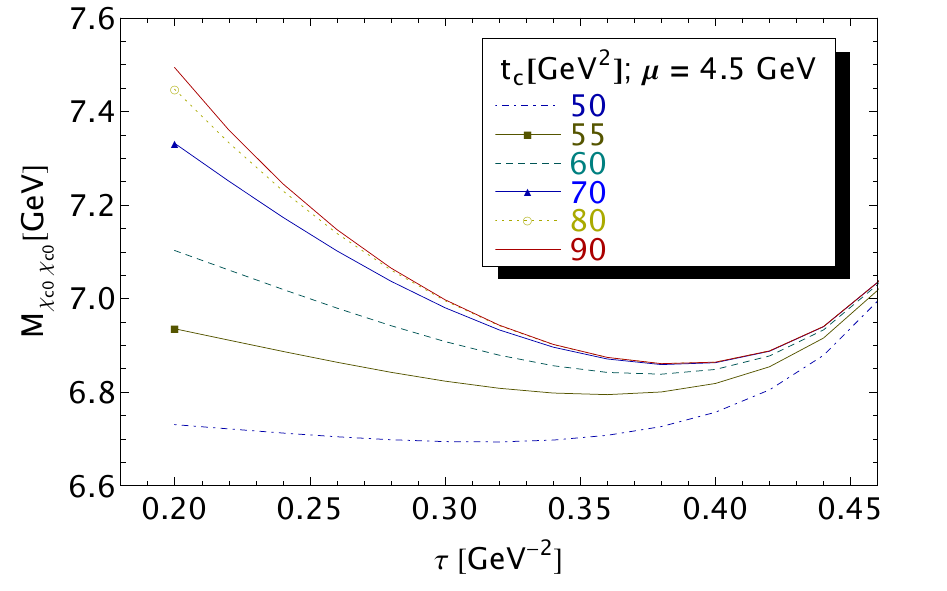}
\vspace*{-0.5cm}
\caption{\footnotesize  $f_{{{\chi}_{c0}}\chi_{c0}}$ and $M_{{{\chi}_{c0}}\chi_{c0}}$ as function of $\tau$ at NLO for different values of $t_c$, for $\mu$=4.5 GeV and for values of $\overline m_{c,b}(\overline m_{c,b})$ given in Table\,\ref{tab:param}.} 
\label{fig:xc-tau}
\end{center}
\vspace*{-0.25cm}
\end{figure}

We show in Fig.\,\ref{fig:xc-tau}, the $\tau$ and $t_c$ behaviours of the $0^{++}({\chi_{c0}}-\chi_{c0})$ molecule fixing $\mu=4.5$ GeV from some other channels,\cite{SNp15,SNbc20,SNbc20b} which we shall justify later.
We see that $f_{{{\chi}_{c0}}\chi_{c0}}$ and  $M_{{{\chi}_{c0}}\chi_{c0}}$ present respectively  inflexion points  and minimas at $\tau\simeq (0.38\pm 0.02)$ GeV$^{-2}$ which appear for $t_c\geq$ 55 GeV$^2$. The $t_c$-stability is reached for $t_c\approx 70$GeV$^2$ We take $t_c\simeq 62.5(7.5)$ GeV$^2$.
\vspace*{-0.25cm}
\subsection{$\mu$-stability}
\begin{figure}[hbt]
\vspace*{-0.25cm}
\begin{center}
\includegraphics[width=9.5cm]{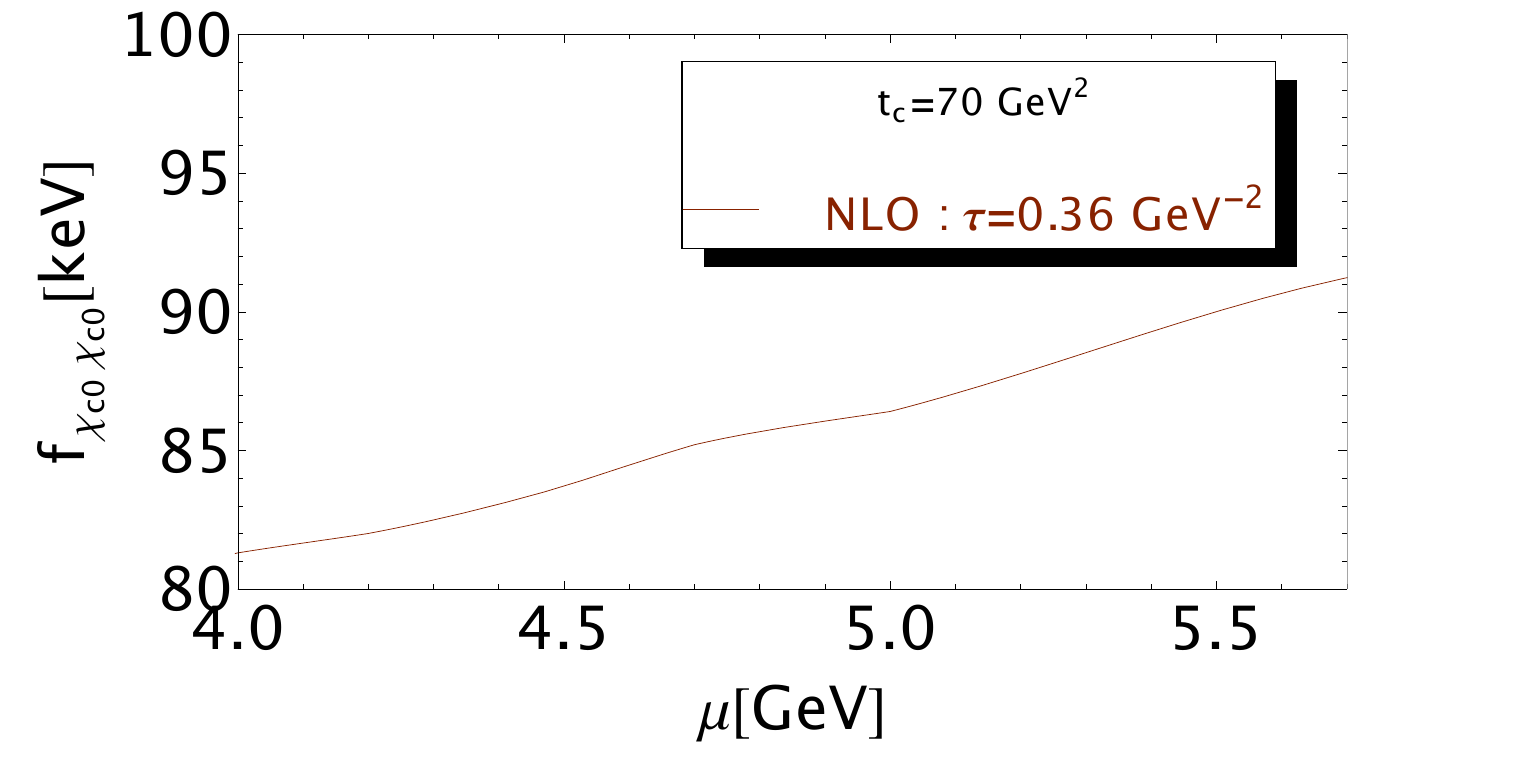}
\includegraphics[width=9.5cm]{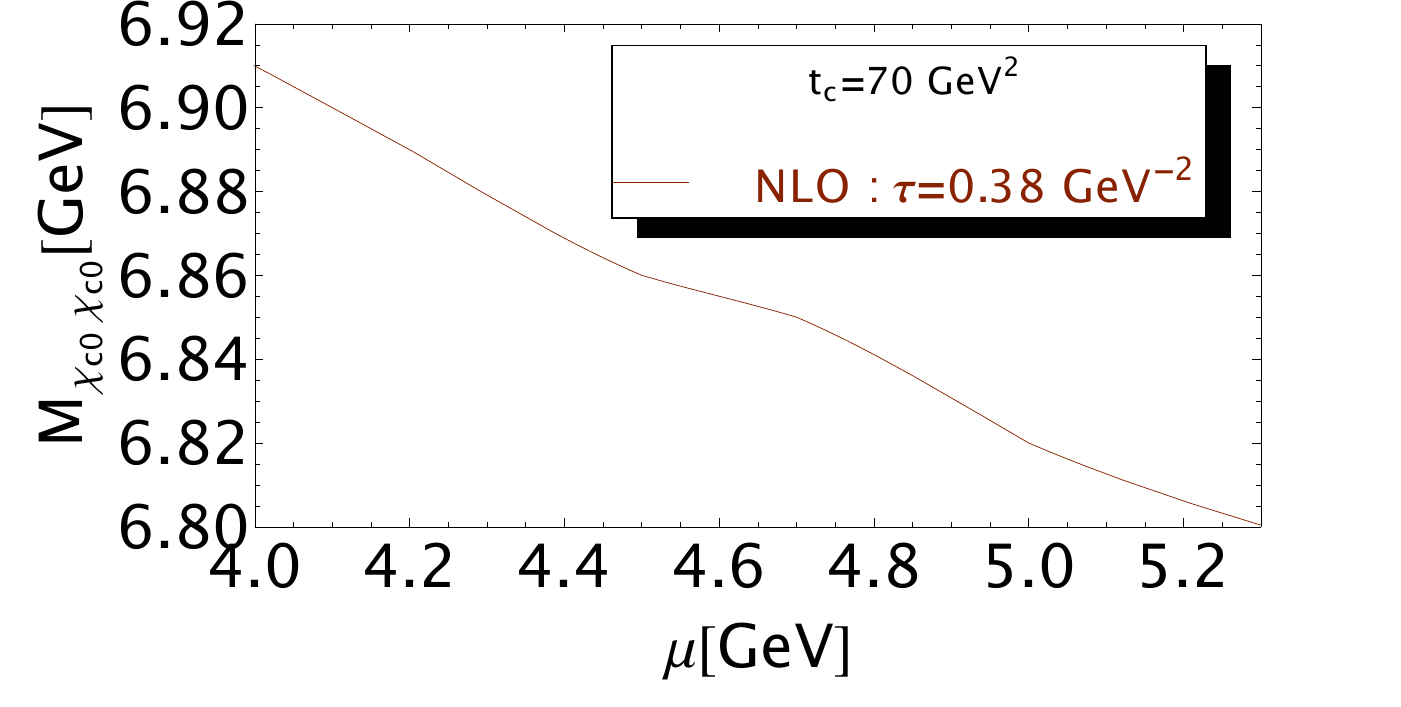}
\vspace*{-0.5cm}
\caption{\footnotesize  $f_{{{\chi}_{c0}}\chi_{c0}}$ and  $M_{{{\chi}_{c0}}\chi_{c0}}$ at NLO as function of $\mu$ for fixed values of $t_c=70$ GeV$^2$ , for $\mu$=4.5 GeV and for values of $\overline m_{c,b}(\overline m_{c,b})$ given in Table\,\ref{tab:param}.} 
\label{fig:xc-mu}
\end{center}
\vspace*{-0.25cm}
\end{figure}

Fixing $t_c=70$ GeV$^2$ and $\tau=(0.35-0.38)$ GeV$^{-2}$, we show in Fig.\,\ref{fig:xc-mu} the $\mu$ behaviour of the mass and coupling where we note an inflexion point at :
\beq
\mu = (4.5\pm 0.2) ~{\rm GeV}~,
\label{eq:mu}
\eeq
in agreement with the one quoted in\,\cite{SNp15,SNbc20,SNbc20b} using different ways and/or from different channels. 

\subsection{QCD continuum versus lowest resonance}
To have more insights on the QCD continuum contribution, we study the ratio of the continuum over the lowest ground state contribution as predicted by QCD :
\beq
r_{{\overline{\chi}_{c0}}\chi_{c0}}\equiv\frac{\int_{t_c}^\infty dt\,{\rm e}^{-t\tau}{\rm Im} \psi_{cont}}{\int_{16m^2}^{t_c}dt\,{\rm e}^{-t\tau}{\rm Im} \psi_{{\overline{\chi}_{c0}}\chi_{c0}}}~.
\label{eq:cont}
\eeq

We found that for $t_c\geq 55$ GeV$^2$, the continuum contribution is less than 60\% of the ground state one and decreases quickly for increasing $t_c$ indicating a complete dominance of the ground state contribution in the sum rule.  
\begin{figure}[hbt]
\vspace*{-0.25cm}
\begin{center}
\includegraphics[width=8cm]{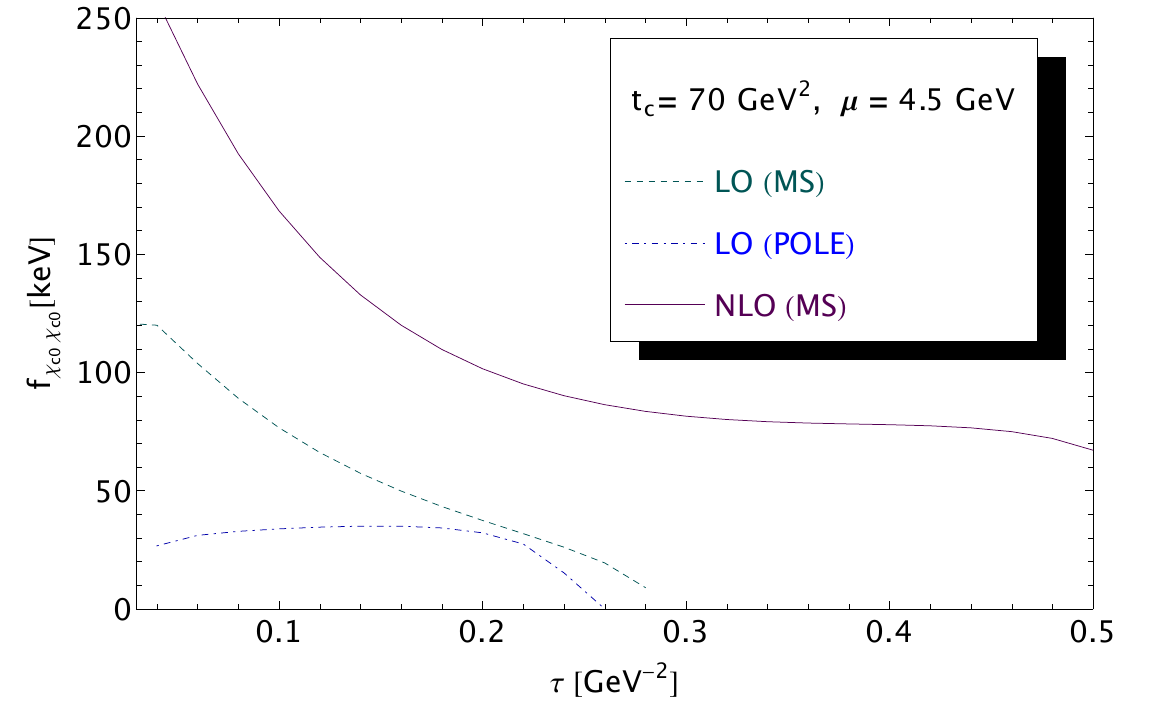}
\includegraphics[width=8cm]{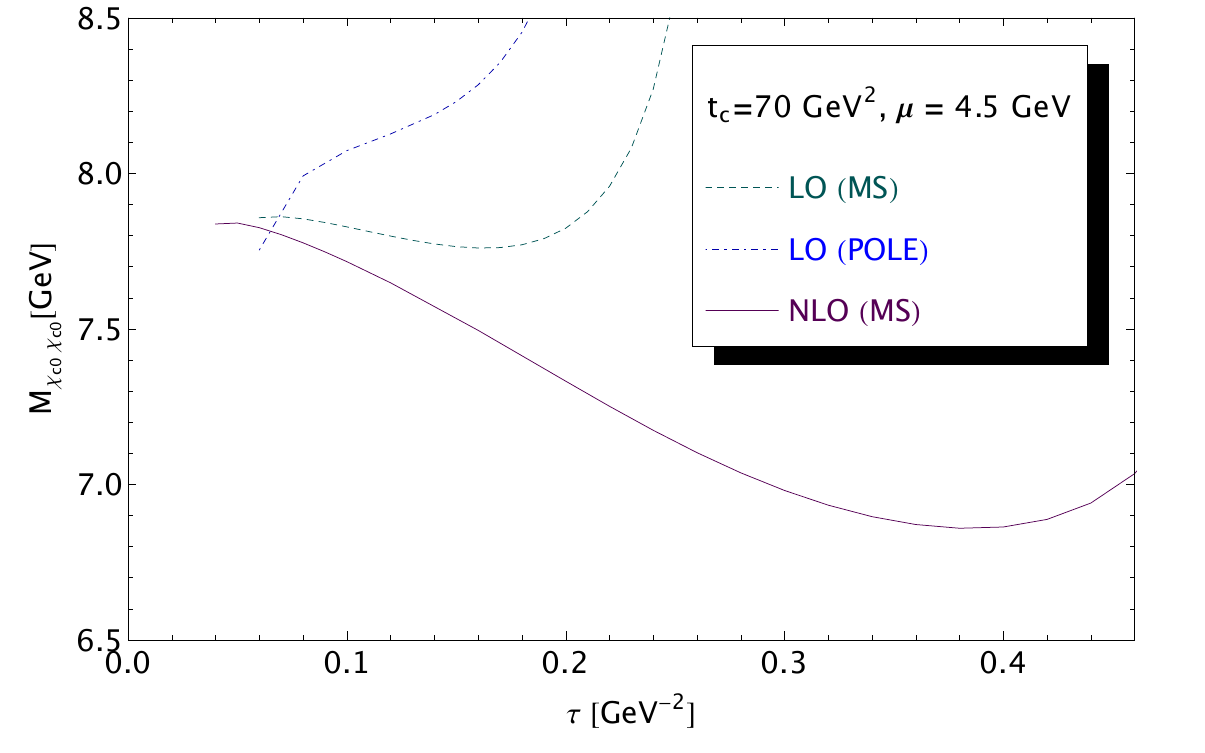}
\vspace*{-0.5cm}
\caption{\footnotesize  Comparison of the LO and NLO contributions on $f_{{{\chi}_{c0}}\chi_{c0}}$ and  $M_{{{\chi}_{c0}}\chi_{c0}}$ as function of $\tau$ for fixed values of $t_c=55$ GeV$^2$ and $\mu$=4.5 GeV.} 
\label{fig:xc-lonlo}
\end{center}
\end{figure}
\subsection{PT series and higher order terms}
   We compare in Fig.\,\ref{fig:xc-lonlo} the LO and NLO perturbative contributions. As the input definition of the quark mass is ambiguous at LO, we use the running mass evaluated at $\mu=4.5$ GeV and the corresponding on-shell / pole mass $M(\mu=M) = 1.53$ GeV. We see that, for the coupling, the two mass definitions lead to about the same predictions but there is a difference about 400 MeV for the mass prediction. This systematic error is never considered
in the literature where a running mass is often used ad hoc with not any justification. This ambiguity is avoided when the PT corrections are added. 

   Comparing the predictions for the running mass at given $\tau\approx 0.17$ GeV$^{-2}$,\,$t_c\simeq 70$ GeV$^2$ and $\mu=4.5$ GeV, one can parametrize numerically the result as :
\bea
\hspace*{-0.5cm}f_{{{\chi}_{c0}}\chi_{c0}}&\approx& 43~{\rm keV} \ga 1+8.7\,a_s \pm 75.7\,a_s^2\dr, \nnb\\
M_{{{\chi}_{c0}}\chi_{c0}}&\approx& 7.76~{\rm GeV}\ga 1-0.5\,a_s \pm 0.25\,a_s^2\dr,
\label{eq:pt-xc}
\eea
where the PT corrections tend to compensate in the ratio of moments used to determine the mass of the meson. We have estimated the N2LO contributions from a geometric growth of the PT coefficients\,\cite{SZ} which we consider as an estimate of the uncalculated higher order terms of the PT series.  

   One can notice, like in the case of  the two-point functions of the scalar quark bilinear currents, that the coefficients of radiative corrections are large for the decay constant\,\cite{SNB1,SNB2,BECCHI}. However,  the PT series converge numerically at $\mu=4.5$ GeV but induce a relatively large systematic error when the higher order terms of the PT series are estimated using a geometric growth of the numerical coefficients. 
\begin{figure}[hbt]
\vspace*{-0.25cm}
\begin{center}
\includegraphics[width=8cm]{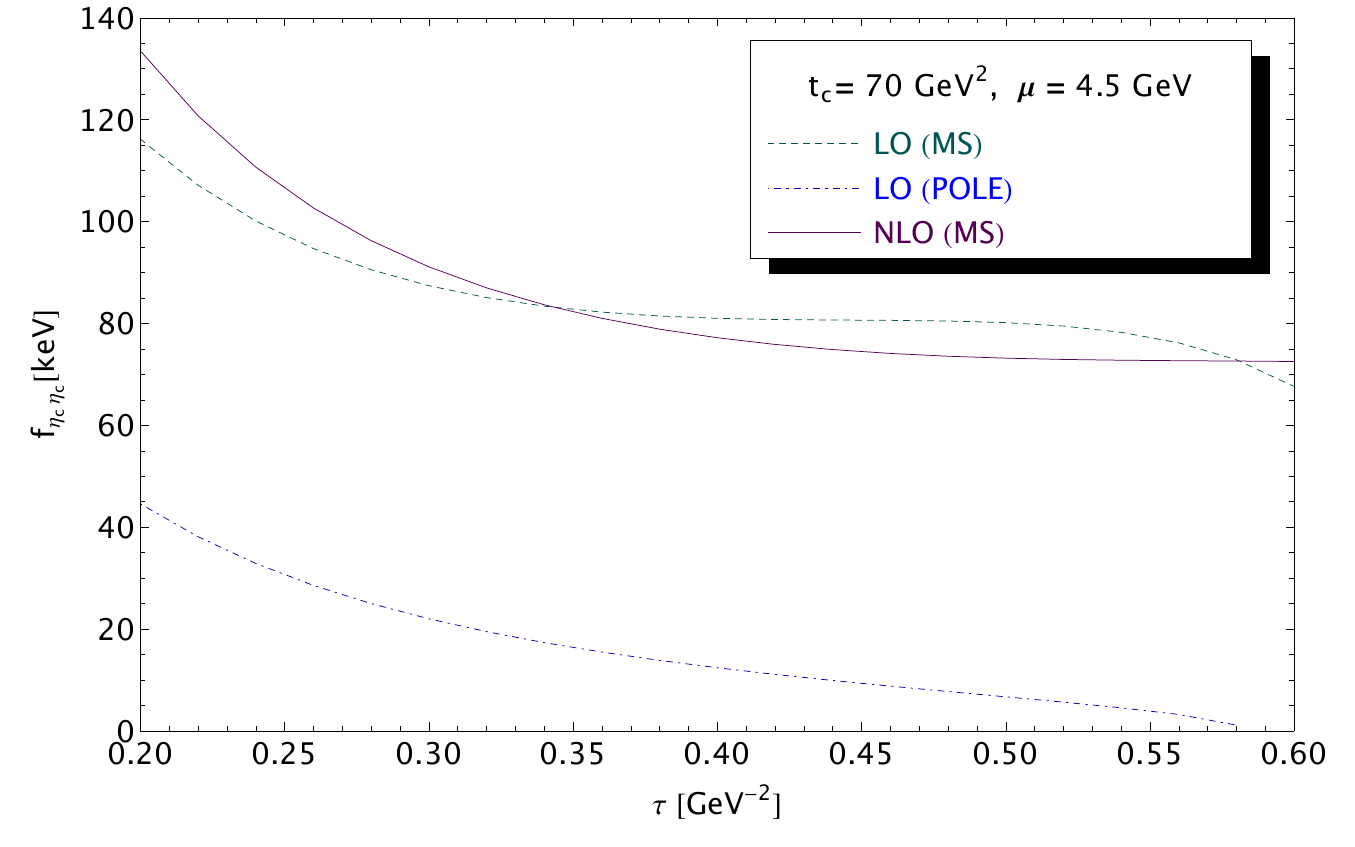}
\includegraphics[width=8cm]{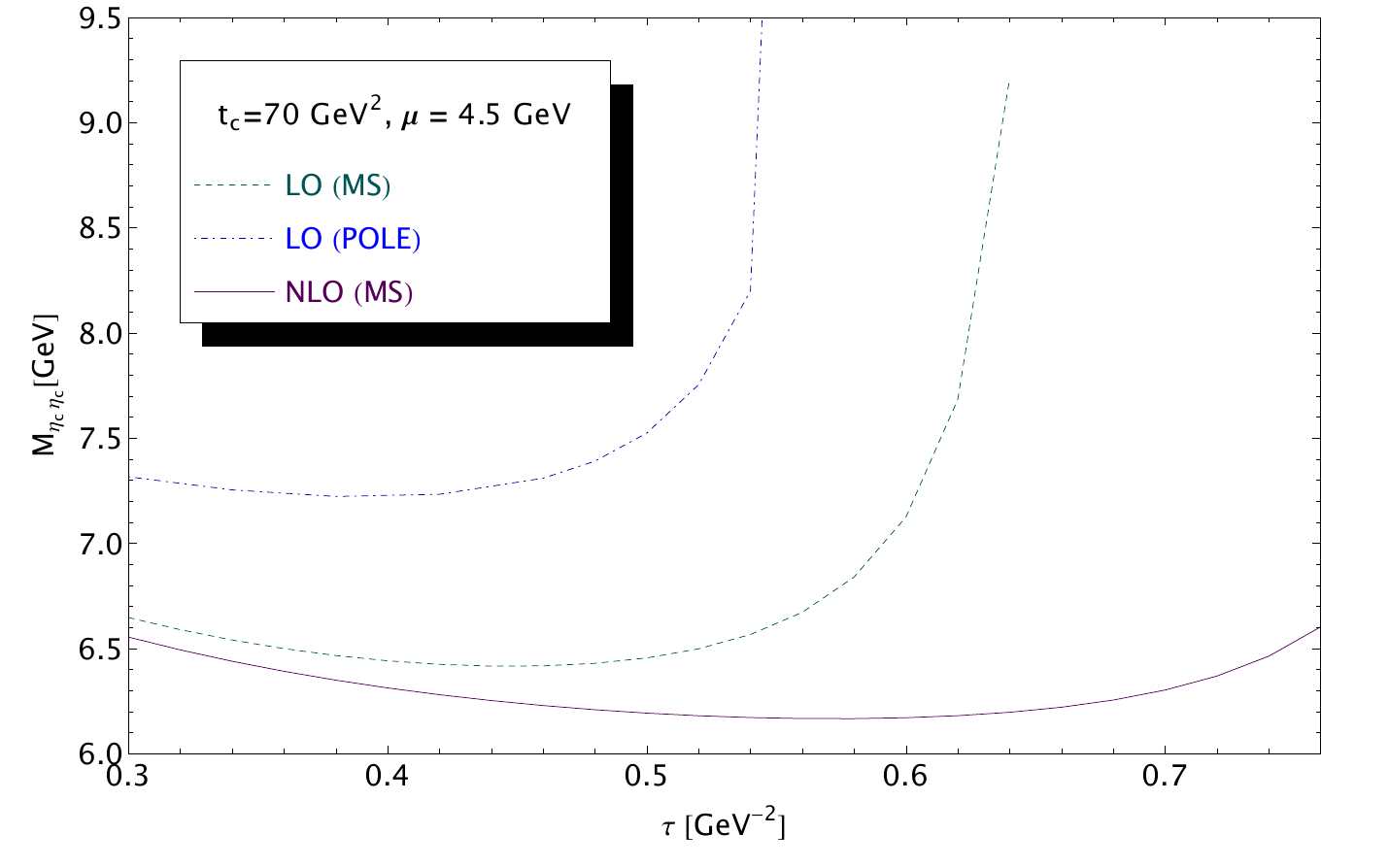}
\vspace*{-0.5cm}
\caption{\footnotesize  Comparison of the LO and LO $\oplus$ NLO contributions on $f_{\overline{\eta}_c\eta_c}$ and  $M_{\overline{\eta}_c\eta_c}$ as function of $\tau$ for fixed values of $t_c=55$ GeV$^2$ and $\mu$=4.5 GeV.} 
\label{fig:etac-lonlo}
\end{center}
\vspace*{-0.25cm}
\end{figure}

\begin{figure}[hbt]
\vspace*{-0.25cm}
\begin{center}
\includegraphics[width=8cm]{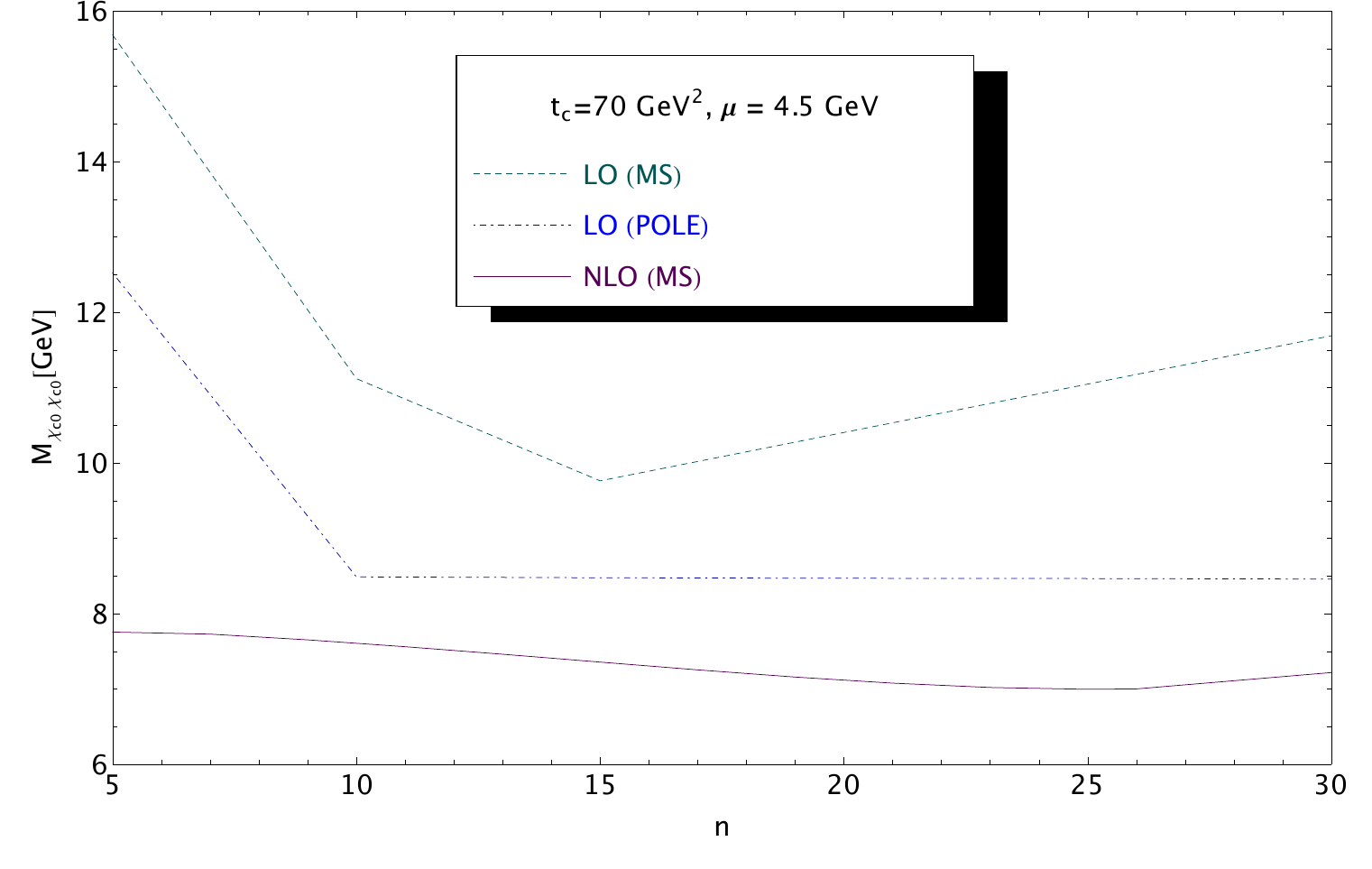}
\vspace*{-0.5cm}
\caption{\footnotesize  Comparison of the LO and LO $\oplus$ NLO contributions on   $M_{\overline{\chi}_{c0}\chi_{c0}}$ as function of the degree $n$ of moments ${\cal M}_n$ for fixed values of $t_c=70$ GeV$^2$ and $\mu$=4.5 GeV.} 
\label{fig:xc-mom}
\end{center}
\vspace*{-0.25cm}
\end{figure}
\section{The ${{\overline{\eta}_{c}}\eta_{c}}$, ${{\overline{J/\psi}}J/\psi}$, $\overline{\chi}_{1c}\chi_{1c}$ molecules }
   The $\tau$, $t_c$ and $\mu$ behaviours of the coupling and mass of these molecules are very similar to the one of $\overline{\chi}_{0c}\chi_{0c}$ and will not be repeated here. The values $\tau$- and $t_c$ at the stability regions are shown in Table\,\ref{tab:lsr-param} where one can notice that, for the  ${{\overline{\eta}_{c}}\eta_{c}}$, the stabilities are reached at earlier values of $t_c$ which is dual to the lower value of the $\overline{\eta}_c\eta_c$ molecule mass. 

   In all cases, the inclusion of the $\la G^3\ra$ condensate shift the $\tau$-stabilty to smaller values. In the case of  the ${{\overline{\eta}_{c}}\eta_{c}}$, it becomes 0.36 GeV$^{-2}$ for the coupling (minimum) and 0.34 GeV$^{-2}$ for the mass (inflexion point).

   The main difference with the $\overline{\chi}_{0c}\chi_{0c}$ as shown in Figs.\ref{fig:xc-lonlo} is the almost equal position of the $\tau$ minima for the  LO and LO $\oplus$ NLO contributions as shown in Fig.\,\ref{fig:etac-lonlo}, which can be attributed to the different reorganisation of the terms in each channel.  

   Our results also emphasize the importance to add radiative PT corrections for a proper heavy quark input (pole or $\overline{MS}$ running) mass definition. In the $\overline{MS}$ scheme, the $\alpha_s$ correction is small as can be seen explicitly in this numerical parametrization :
\bea
f_{\overline{\eta}_c\eta_c}&\simeq& 80~{\rm keV} \ga 1-1.4\,a_s \pm 1.96\,a_s^2\dr,\nnb\\
M_{\overline{\eta}_c\eta_c}&\simeq& 6.4~{\rm GeV}\ga 1-0.57\,a_s \pm 0.32\,a_s^2\dr~.
\label{eq:pt-etac}
\eea
    The $\mu$-stability is reached at $\mu=$ 4.5 GeV. The results of the analysis are shown in Table\,\ref{tab:res}. 
 
\section{The ${{\overline{\chi}_{b0}}\chi_{b0}}$  molecule}
The extension of the analysis to the $b$ quark channel is straigthforward.
We show in this example the details of the analysis.
\subsection{$\tau$- and $t_c$-stabilities}
\begin{figure}[hbt]
\vspace*{-0.25cm}
\begin{center}
\includegraphics[width=8cm]{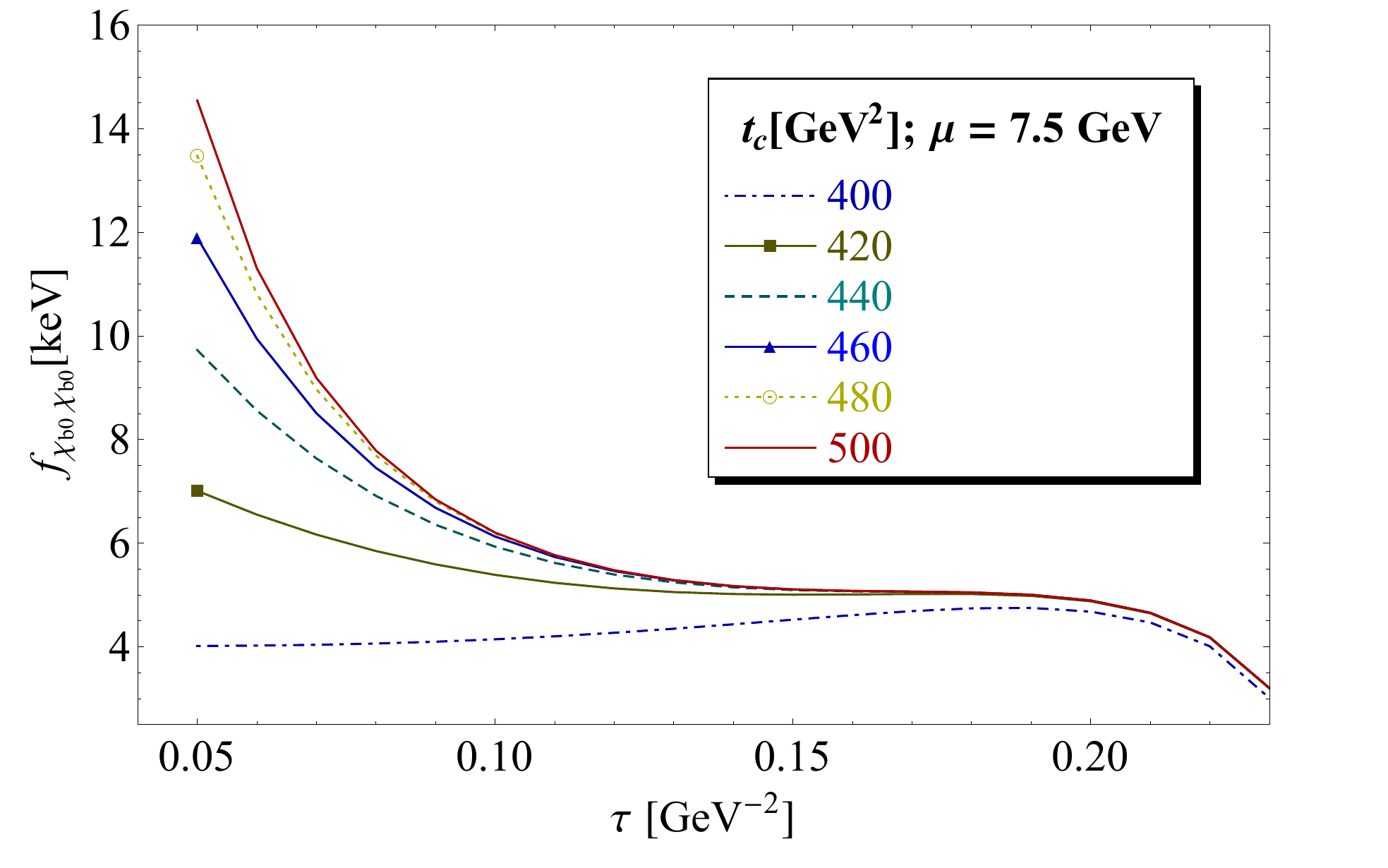}
\includegraphics[width=8cm]{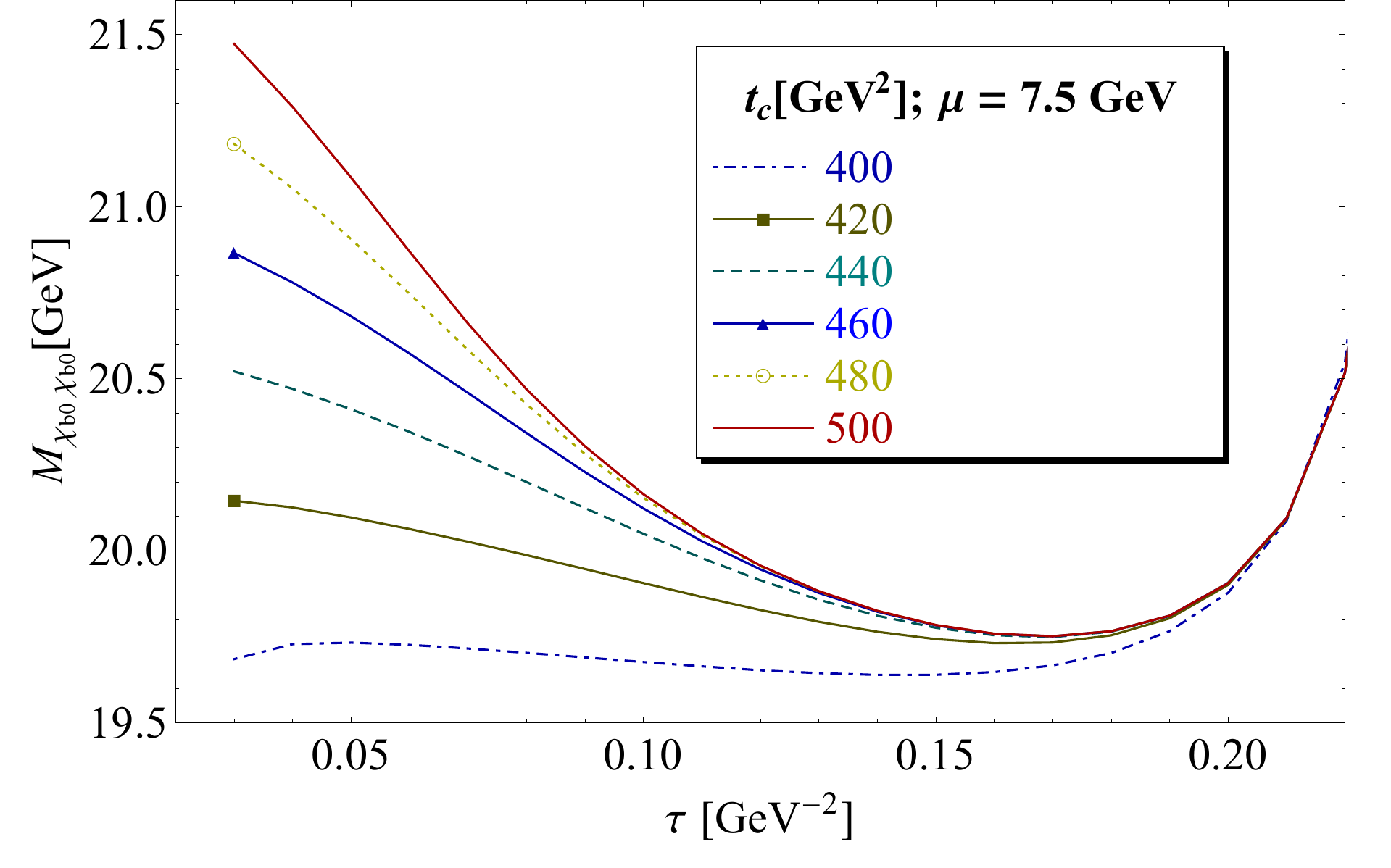}
\vspace*{-0.5cm}
\caption{\footnotesize  $f_{{{\chi}_{c0}}\chi_{b0}}$ and $M_{{{\chi}_{b0}}\chi_{b0}}$ as function of $\tau$ at NLO for different values of $t_c$, for $\mu$=7.5 GeV and for values of $\overline m_{b}(\overline m_{b})$ given in Table\,\ref{tab:param}.} 
\label{fig:xb-tau}
\end{center}
\vspace*{-0.25cm}
\end{figure}

  The $\tau$ and $t_c$ behaviours of the $0^{++}(\overline{\chi}_{b0}\chi_{b0})$ molecule fixing $\mu=7.5$ GeV from some other channels\,\cite{SNp15,SNbc20,SNbc20b} are shown in Fig.\,\ref{fig:xb-tau}, where the stability (minimas and inflexion points)  is reached for $\tau\simeq 0.17$ GeV$^{-2}$  and $t_c\simeq (420-460)$ GeV$^2$. 

   The $\mu$-stabilty is shown in Fig.\,\ref{fig:xb-mu}.

\vspace*{-0.25cm}
\subsection{$\mu$-stability}
\begin{figure}[hbt]
\vspace*{-0.25cm}
\begin{center}
\includegraphics[width=9.5cm]{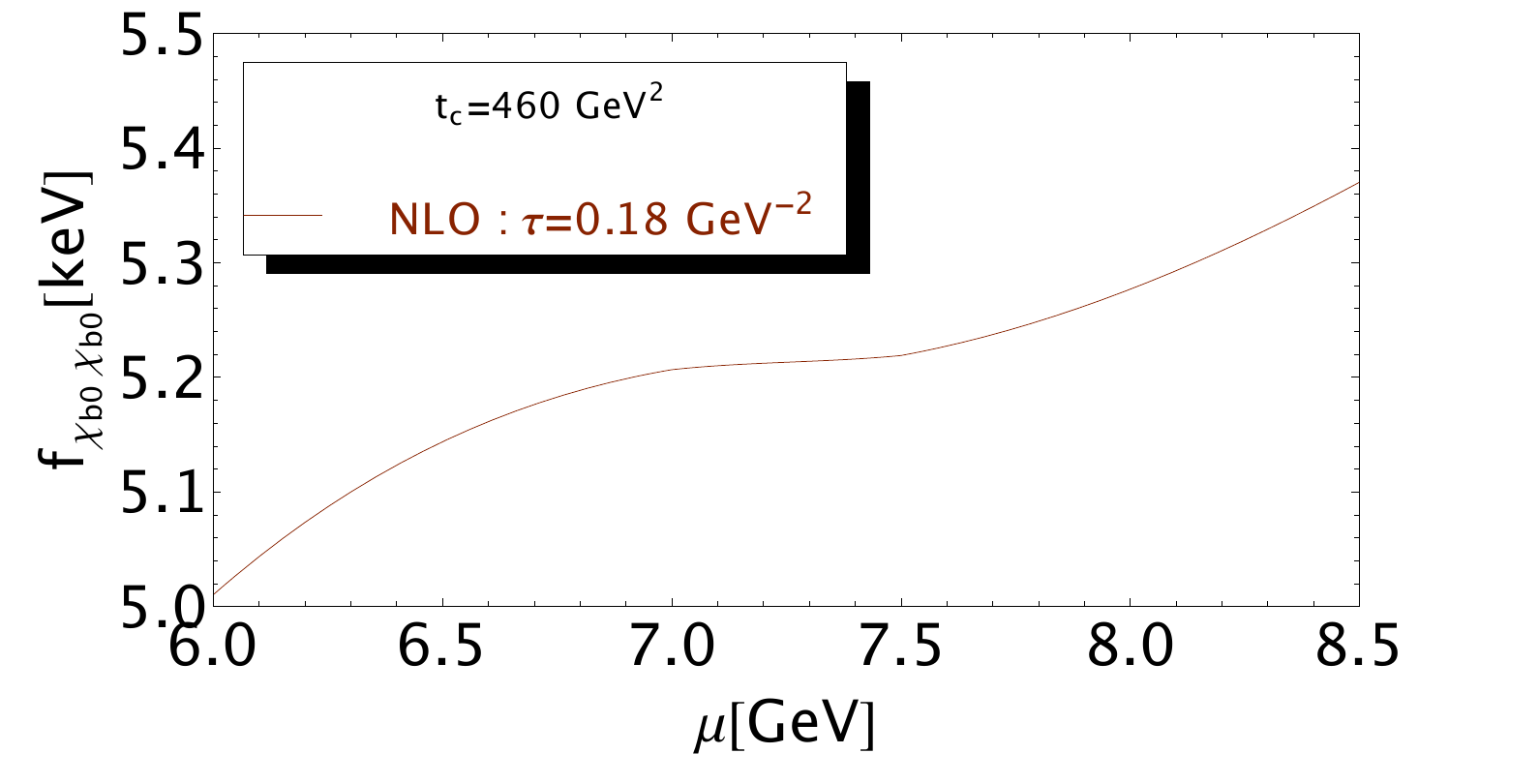}
\includegraphics[width=9.5cm]{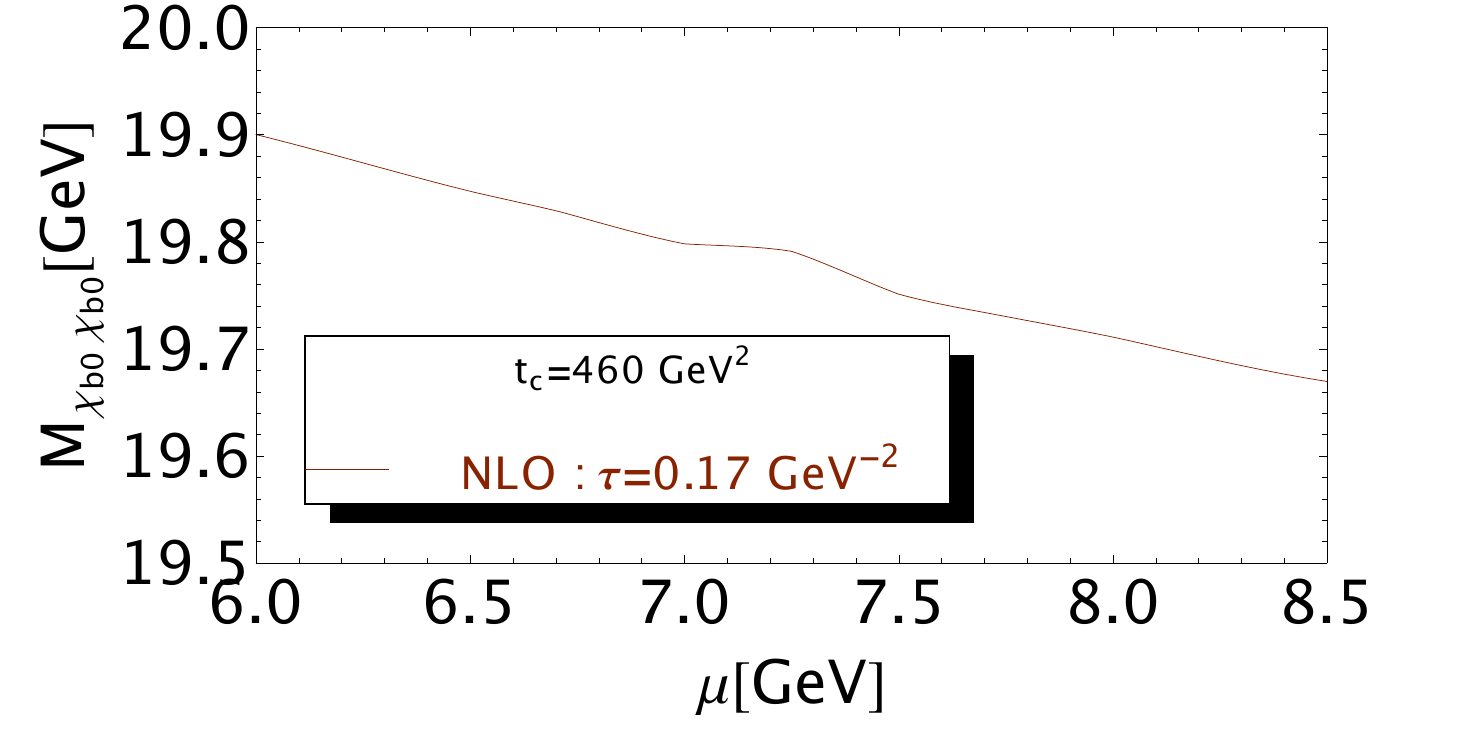}
\vspace*{-0.5cm}
\caption{\footnotesize  $f_{{\chi}_{b0}\chi_{b0}}$ and  $M_{{{\chi}_{b0}}\chi_{b0}}$ at NLO as function of $\mu$ for fixed values of $t_c=460$ GeV$^2$ , for $\mu$=4.5 GeV and for values of $\overline m_{b}(\overline m_{b})$ given in Table\,\ref{tab:param}.} 
\label{fig:xb-mu}
\end{center}
\vspace*{-0.25cm}
\end{figure}

Fixing $t_c=460$ GeV$^2$ and $\tau=0.17$ GeV$^{-2}$, we show in Fig.\,\ref{fig:xb-mu} the $\mu$ behaviour of the mass and coupling, where we find a clear inflexion point for the coupling  but a slight for the mass at :
\beq
\mu = (7.25\pm 0.25) ~{\rm GeV}~,
\label{eq:mub}
\eeq
in agreement with the one quoted in\,\cite{SNp15,SNbc20,SNbc20b} using different ways and/or from different channels. 
\vspace*{-0.25cm}
\subsection{LO versus NLO contributions}
\begin{figure}[hbt]
\vspace*{-0.25cm}
\begin{center}
\includegraphics[width=8cm]{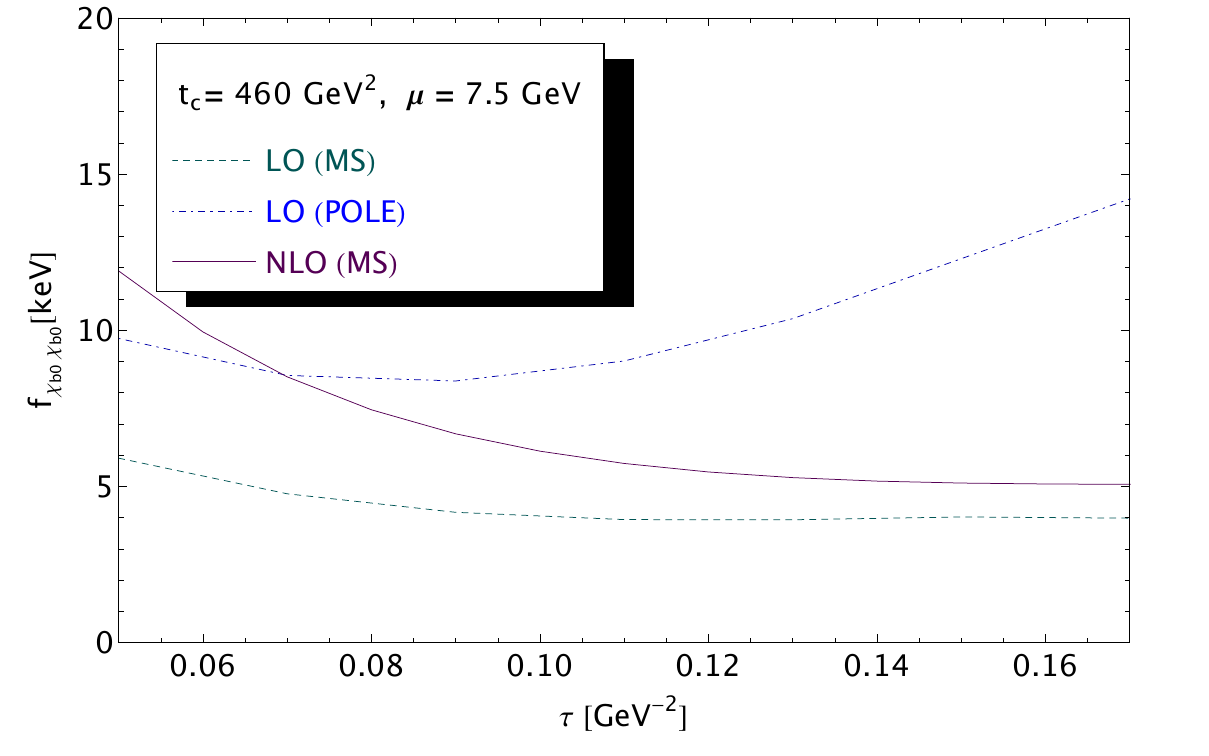}
\includegraphics[width=8cm]{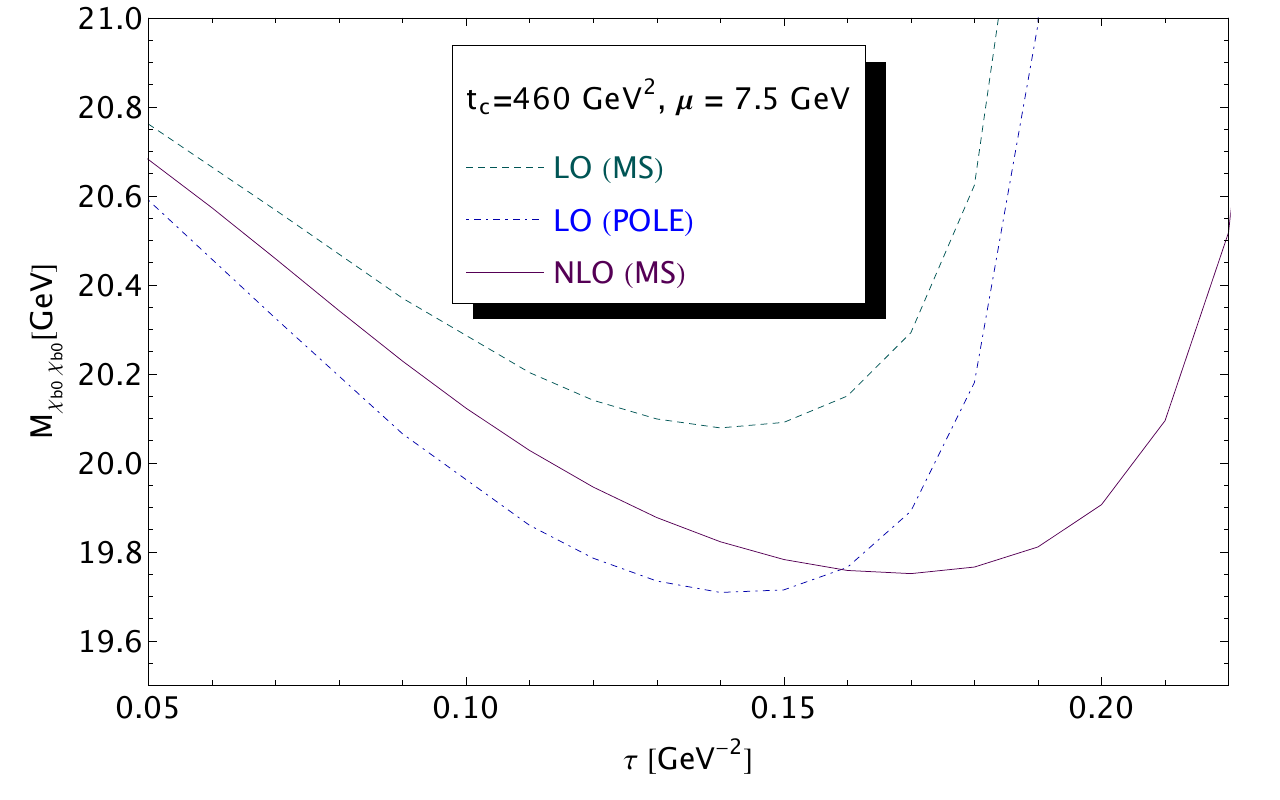}
\vspace*{-0.5cm}
\caption{\footnotesize  Comparison of the LO and LO $\oplus$ NLO contributions on $f_{{\overline{\chi}_{b0}}\chi_{b0}}$ and  $M_{{\overline{\chi}_{b0}}\chi_{b0}}$ as function of $\tau$ for fixed values of $t_c=460$ GeV$^2$ and $\mu$=7.5 GeV.} 
\label{fig:xb-lonlo}
\end{center}
\vspace*{-0.25cm}
\end{figure}
We compare in Fig\,\ref{fig:xb-lonlo} the LO and LO $\oplus$ NLO contributions. We note (as expected) that the radiative corrections is smaller for $b$ than for $c$ as the coupling and mass are evaluated at higher $\mu$-values. Using this result, we can numerically parametrize the previous observables as:
\bea
f_{{{\chi}_{b0}}\chi_{b0}}&\simeq& 3.9~{\rm keV} \ga 1+3.8\,a_s \pm 14.4\,a_s^2\dr \nnb\\
M_{{{\chi}_{b0}}\chi_{b0}}&\simeq& 20.1~{\rm GeV}\ga 1-0.3\,a_s \pm 0.1\,a_s^2\dr
\label{eq:pt-xb}
\eea
where the PT corrections tend to compensate in the ratio of moments while, compared to the $c$-quark channel, the PT corrections are  relatively small. As in the previous cases, we have estimated the N2LO contributions from a geometric growth of the PT coefficients\,\cite{SZ} which we consider as an estimate of the uncalculated higher order terms of the PT series.  
\begin{figure}[hbt]
\vspace*{-0.25cm}
\begin{center}
\includegraphics[width=8cm]{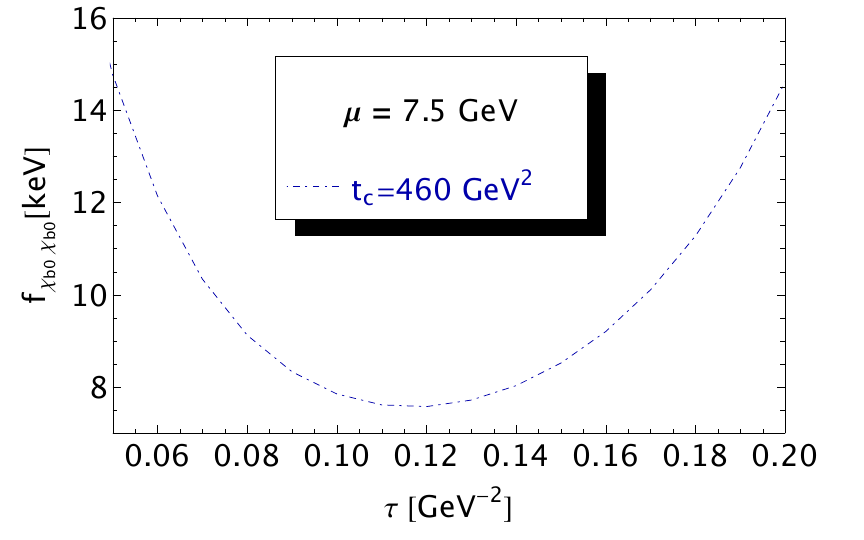}
\includegraphics[width=8cm]{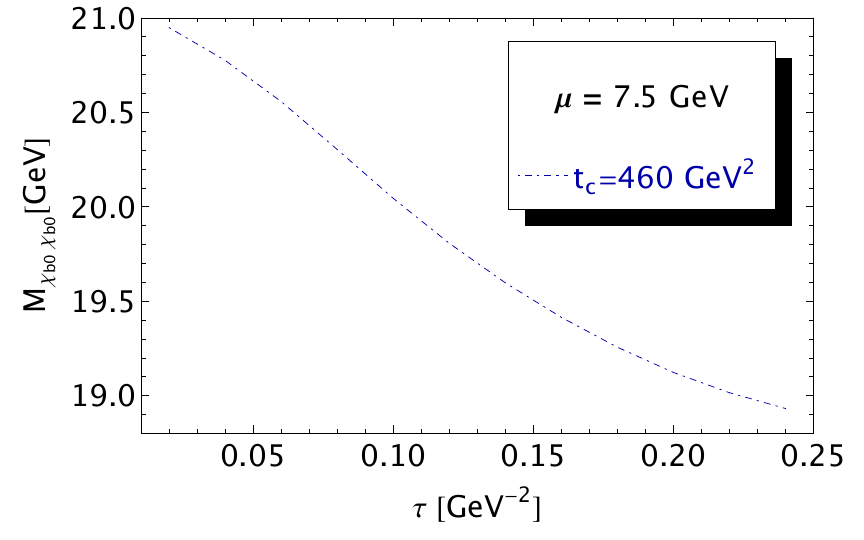}
\vspace*{-0.5cm}
\caption{\footnotesize  Effect of the $\la G^3\ra$ condensate on the $\tau$-behaviour of  $f_{{\overline{\chi}_{b0}}\chi_{b0}}$ and  $M_{{\overline{\chi}_{b0}}\chi_{b0}}$ for fixed values of $t_c=460$ GeV$^2$ and $\mu$=7.5 GeV.} 
\label{fig:xbg3}
\end{center}
\vspace*{-0.25cm}
\end{figure}
\vspace*{-0.25cm}
\section{$\la G^3\ra$ and truncation of the OPE}\label{sec:OPE}
   We have included the $\la G^3\ra$ condensate contribution into the sum rule. We have cross-checked that with our method of calculation we reproduce the results of\,\cite{BAGAN1} for charmonium sum rules.

   We have noticed that in  the $\overline{\chi}_{c0}\chi_{c0}$ channel, the contribution of the $\la G^3\ra$ condensate is relatively small and does not modify the shape of the mass and coupling curves versus the variation of $\tau$ and for different values of $t_c$. It only decreases the decay constant by 0.4 keV and increases the mass by 14 MeV. 

   However, this is not the case of some other channels which will be analyzed later on where the $\la G^3\ra$ contribution can be large and modify the minimum of the mass found for $\la \alpha_sG^2\ra$ into an inflexion point (see Fig.\,\ref{fig:xbg3}) and vice-versa for the coupling.  This feature renders the mass result quite sensitive to the localisation of this inflexion point.  An analogous effect of  $\la G^3\ra$ has been also observed e.g in the analysis of charmonium sum rules\,\cite{SNH10,SNH11,SNH12} and  the inclusion of the $\la G^4\ra$ condensates which act with an opposite sign restores the stability of these sum rules.

   To circumvent this problem and due to the difficulty for evaluating  the $\la G^4\ra$ contribution, we consider the optimal result at the value of $\tau$ where the coupling presents a minimum. Then we consider as a final result (here and in the following), the mean obtained with and without the $\la G^3\ra$ contribution. The error induced in this way will be included as the systematics due to the truncation of the OPE as quoted in Table\ref{tab:res}. 
\section{The ${{\overline{\eta}_{b}}\eta_{b}}$ ,  ${{\overline{\Upsilon}}\Upsilon}$, $\overline{\chi}_{1b}\chi_{1b}$ molecules}
The analysis of these scalar molecules is very similar to the analysis presented above. 
The QCD expressions of their corresponding two-point functions are given in Appendix A. One should mention that in these channels the PT radiative corrections and the contribution of the $\la G^3\ra$ condensate are small indicating a good convergence of the PT series and of the OPE at the optimization scale. The results are quoted in Table\,\ref{tab:res} where the LSR parameters used to get them are shown in Table\,\ref{tab:lsr-param}. 

\section{The scalar tetraquark states}
   We repeat the previous analysis for the case of tetraquark states with same choice of diquark currents as in\,\cite{STEELE} :
\beq
J^{[P,S,A,V]}_{\cal T}=Q^T_aC[1,\gamma_5,\gamma_\mu, \gamma_5\gamma_\mu]Q_b~,
\label{eq:tetra-steele}
\eeq
in order to make a direct comparison with their LO results. We do not consider the current associated to $\sigma_{\mu\nu}$ which corresponds to a two-point correlator of higher dimension. We shall also consider the four-quark operator :
\beq
{\cal O}_{\cal T}=\epsilon_{abc}\epsilon_{cde}(Q^T_aC\gamma_\mu Q_b)(Q^T_dC\gamma^\mu Q_e)~,
\label{eq:tetra-wang}
\eeq
in order to make a direct comparison with\,\cite{WANG}. One should notice that due to the epsilon-tensor, most of the currents used by\,\cite{STEELE} are not present in\,\cite{WANG}. 

   The QCD expressions of their corresponding two-point functions are given in Appendix B.

   The behaviours of different curves are very similar with the ones of the corresponding molecule case. 

   We quote the results in Table\,\ref{tab:res} and the optimal LSR parameters used to get them in Table\,\ref{tab:lsr-param}. These results are compared with the ones in\,\cite{STEELE,WANG} in Table\,\ref{tab:comparison}.
\vspace*{-0.25cm}
\begin{table*}[hbt]
\setlength{\tabcolsep}{0.25pc}
\catcode`?=\active \def?{\kern\digitwidth}
    {\footnotesize
  \begin{tabular*}{\textwidth}{@{}l@{\extracolsep{\fill}}|ccccccccc   c cccccc   ll}
\hline
\hline
Observables\, 
 &  \multicolumn{2}{c}{$ \Delta t_c$} 
  &  \multicolumn{2}{c}{$\Delta\tau$} 
     &  \multicolumn{2}{c}{$\Delta\mu$} 
       &  \multicolumn{2}{c}{$\Delta m$} 
         &  \multicolumn{2}{c}{$\Delta\alpha_s$} 
         &  \multicolumn{2}{c}{$\Delta\alpha_s G^2$} 
         &  \multicolumn{2}{c}{$\Delta G^3$-OPE} 
         &  \multicolumn{2}{c}{HO-PT} 
           &  \multicolumn{2}{c}{Values} 
          \\

 \cline{0-1}  \cline{2-3}\cline{4-5}\cline{6-7}\cline{8-9}\cline{10-11}
   \cline{12-13}\cline{14-15} \cline{16-17}\cline{18-19}
  
            $q\equiv c,b$& \multicolumn{1}{c}{c} & \multicolumn{1}{c}{b} 
          & \multicolumn{1}{c}{c} & \multicolumn{1}{c}{b} 
           & \multicolumn{1}{c}{c} & \multicolumn{1}{c}{b} 
            & \multicolumn{1}{c}{c} & \multicolumn{1}{c}{b} 
             & \multicolumn{1}{c}{c} & \multicolumn{1}{c}{b} 
              & \multicolumn{1}{c}{c} & \multicolumn{1}{c}{b} 
               & \multicolumn{1}{c}{c} & \multicolumn{1}{c}{b} 
                & \multicolumn{1}{c}{c} & \multicolumn{1}{c}{b} 
                 & \multicolumn{1}{c}{c} & \multicolumn{1}{c}{b} 

                 \\


 \cline{0-1}  \cline{2-3}\cline{4-5}\cline{6-7}\cline{8-9}\cline{10-11}
   \cline{12-13}\cline{14-15} \cline{16-17}\cline{18-19}
$f_H$ [keV] \\
\cline{0-0} 
\footnotesize $0^{++}$ Molecule \\
\cline{0-0}
${\overline\eta_{q}\eta_{q}}$&0.8&0.4&0.2&0.1&3.0&0.2&10&1.2&5&2&0.7&0.1&12.2&0.8&0.9&0.2&$56\pm 17$&$9.8\pm 2.4$\\
${\overline{J/\psi} J/\psi,\overline\Upsilon\Upsilon}$&4.6&0.6&1.0&0.6&2.0&0.1&10.7&4.3&19&2.5&3.4&0.4&45.6&3.8&0.4&0&$160\pm 51$&$23.4\pm 6.3$\\
${\overline\chi_{q1}\chi_{q1}}$&0.9&1.6&1.1&0.9&0.9&0.2&6&3&9&4.8&10&0.0&4&3&5&19&$162\pm 16$&$48.9\pm 20.1$ \\
${\overline\chi_{q0}\chi_{q0}}$
&2.8&0.01&0.4&0.1&2.5&0.1&3.7&0.5&3.5&0.7&1.2&0.1&11.5&0.6&16&0.2&$69\pm 21$&$4.0\pm1.1$\\
\\
\cline{0-0}
\footnotesize $0^{++}$ Tetraquark\\
\cline{0-0}
\footnotesize\bf Eq.\,\ref{eq:tetra-steele}\\
${\overline S_qS_q}$&0.1&0.1&0.7&0.2&9&2.3&20&2.3&9&3.7&0.3&0.1&7&9&87&0.1&$249\pm 90$&$29.6\pm 10.2$\\
${\bar A_qA_q} $&1.4&4.1&1.0&7.2&1.5&3.4&19.2&4.0&8.8&6.4&0.36&0.&10&2.8&65&27&$220\pm 69$&$87.4\pm 29.5$ \\
${\overline V_qV_q} $&5.2&0.4&1.0&0.3&6.5&0.3&11.8&1.5&5.4&2.4&1.9&0.2&9&0.3&0.9&0.1&$102\pm 18$&$17.2\pm 2.9$ \\
${\overline P_qP_q}$&1.4&1.8&0.4&2.3&3.4&0.5&7.2&1&3.5&1&1.3&0.1&8.9&3.5&4.8&1.2&$60\pm 14$&$6.5\pm 4.9$ \\
\footnotesize \bf Eq.\,\ref{eq:tetra-wang}\\
${\bar A_qA_q} $&3&3.6&1.5&2&4.8&2&37.5&7.7&17.6&12.3&0.8&0.1&12&7&108&72 &$448\pm 117$&$136\pm 74$ \\
\\
\hline
$M_H$ [MeV]\\
\cline{0-0}
\footnotesize $0^{++}$ Molecule \\
\cline{0-0}
${\overline\eta_{q}\eta_{q}}$&23&4&3&15 &23&26&51&29&24&49&14&13&186&58&3.8&1.6&$6029\pm 198$&$19259\pm 88$ \\
${\overline{J/\psi }J/\psi,\overline\Upsilon\Upsilon}$&34&31&11&42&24&27&27&52&49&30&31&22&359&116&1.3&0&$6376\pm 367$&$19430\pm 145$ \\
${\overline\chi_{q1}\chi_{q1}} $&26&4&29&99&20&22&42&25&20&43&5&22&16&73&7&6&$6494\pm 66$&$19770\pm 137$\\
${\overline\chi_{q0}\chi_{q0}} 
$ &11&39&8&28&10&24&47&36&19 &18&29&13&76&112&9&8&$6675\pm 98$&$19653\pm131$\\
\\
\cline{0-0}
\footnotesize $0^{++}$ Tetraquark\\
\cline{0-0}
\footnotesize \bf Eq.\,\ref{eq:tetra-steele}\\
${\overline S_qS_q} $ &12&1&28&38&21&26&54&29&43&59&1&2&25&89&9&9&$6411\pm 83$&$19217\pm 120$\\
${\bar A_qA_q} $&26&37&32&132&20&23&43&25&21&43&2&1&38&53&0.0&10&$6450\pm 75$&$19872\pm156$ \\
${\overline  V_qV_q} $&59&27&10&22&26&4&47&29&25&50&21&15&152&39&1&0.1&$6462\pm 175$&$19489\pm 79$ \\
${\overline P_qP_q}$&34&10&19&40&23&24&46&28&20&46&30&22&258&23&22&5&$6795\pm 268$&$19754\pm 79$ \\
\footnotesize \bf Eq.\,\ref{eq:tetra-wang}\\
${\bar A_qA_q} $&4&21&3&95&21&25&43&27&21&47&2&0&39&30&16&2&$6471\pm 67$&$19717\pm 118$\\
\hline\hline
\end{tabular*}
{\scriptsize
 \caption{Predictions from LSR at NLO and sources of errors for the decay constants and masses of the molecules and tetraquark states. The errors from the QCD input parameters are from Table\,\ref{tab:param}. $\Delta \mu$ are given in Eqs.\,\ref{eq:mu} and\,\ref{eq:mub}. We take $|\Delta \tau|= 0.02$ GeV$^{-2}$.
 In the case of asymetric errors, we take the mean value. The inclusion of the $\la G^3\ra$ contribution and the way to estimate the systematics induced by the truncation of the OPE are explained in Section\,\ref{sec:OPE}.
 }
 \label{tab:res}
}
}
\end{table*}


\begin{table*}[hbt]
\setlength{\tabcolsep}{0.pc}
\catcode`?=\active \def?{\kern\digitwidth}
 {\footnotesize
\begin{tabular*}{\textwidth}{@{}l@{\extracolsep{\fill}}|cccccc  cc | cccc cccc}
\hline
\hline
      Scalar          & \multicolumn{8}{c |}{Molecules} 
                  & \multicolumn{8}{ c }{Tetraquarks} \\
\cline{1-9} \cline{10-17}
Parameters          & \multicolumn{1}{c}{$\overline{\eta}_c\eta_c$} 
       & \multicolumn{1}{c}{$\overline{J/\psi}J/\psi$} 
                 & \multicolumn{1}{c}{{$~\overline{\chi}_{c0}\chi_{c0}$}} 
                 & \multicolumn{1}{c}{{$~\overline{\chi}_{c1}\chi_{c1}$} } 
              & \multicolumn{1}{c}{$\overline{\eta}_b\eta_b$} 
       & \multicolumn{1}{c}{$\overline{\Upsilon}\Upsilon$} 
                 & \multicolumn{1}{c}{{$\overline{\chi}_{b0}\chi_{b0}$}} 
                 & \multicolumn{1}{c |}{{$\overline{\chi}_{b1}\chi_{b1}$} } 
                 
                  & \multicolumn{1}{ c}{$\overline{S}_cS_c$} 
       & \multicolumn{1}{c}{$\bar{A}_cA_c$}
            & \multicolumn{1}{c}{{$\overline{V}_cV_c$} } 
                 & \multicolumn{1}{c}{{$\overline{P}_c P_c$}} 
                       & \multicolumn{1}{c}{$\overline{S}_bS_b$} 
       & \multicolumn{1}{c}{$\bar{A}_bA_b$}
               & \multicolumn{1}{c}{{$\overline{V}_bV_b$} } 
                 & \multicolumn{1}{c}{{$\overline{P}_bP_b$}} 
                                \\
\cline{1-5} \cline{6-9} \cline{10-13} \cline{14-17}
\\
$t_c$ [GeV$^2$]&45  55&55  70&55  70&55  70&400-460&400  460&420  460&420  460&55  70&55  70& 50  70&60  90&400  460 &420  460&400  460&420  460\\
$\tau$ [GeV$^{-2}$]$10^2$&50, 54&30, 34&36, 38&34&21,22&14, 16&16, 17&7,9&34&32  38&38&32, 34&22&6  8&15, 16&8  18\\

\hline\hline
\end{tabular*}
 \caption{Values of the LSR parameters $t_c$ and the corresponding $\tau$ at the otpimization region for the PT series up to NLO and for the OPE truncated at $\la\alpha_s G^2\ra$.}  
 }
\label{tab:lsr-param}
\end{table*}

\section{Comments on the results}
\subsection{The quest of factorization and Landau singularities}
\begin{figure}[hbt]
\vspace*{-0.25cm}
\begin{center}
\includegraphics[width=8cm]{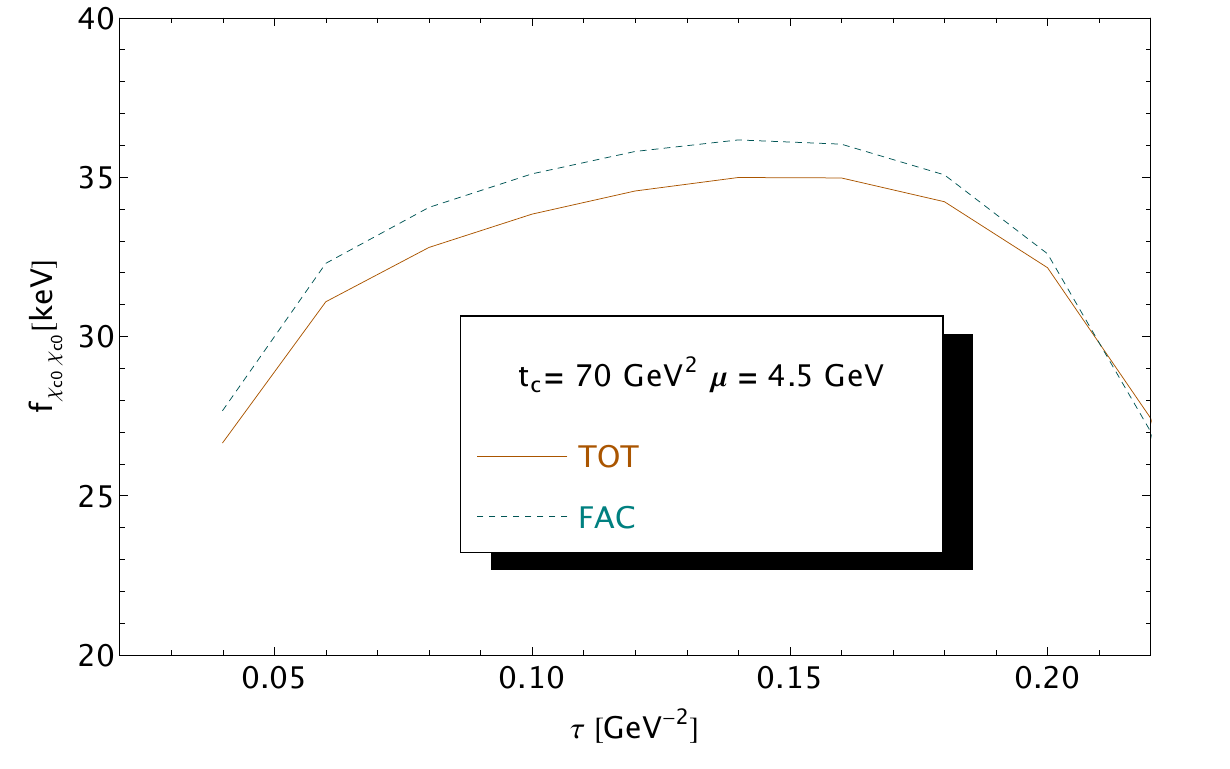}
\vspace*{-0.5cm}
\caption{\footnotesize  Comparison of the factorized and factorized $\oplus$ non-factorized (TOT) at LO including the $\alpha_s G^2$ condensate contribution to the decay constant $f_{{{\chi}_{c0}}\chi_{c0}}$  versus the $\tau$ for fixed values of $t_c=70$ GeV$^2$ and $\mu$=4.5 GeV. We use the pole mass of 1.53 GeV.} 
\label{fig:fac}
\end{center}
\end{figure}

   We have shown explictily in Eq.\,\ref{eq:scalar} that the contributions from the non-factorized diagrams appear already to LO of perturbative series and for the lowest dimension $\la \alpha_sG^2\ra$
gluon condensate contributions. This result does not support the claims of\,\cite{LUCHA1,LUCHA2} that non-factorized contributions start to order $\alpha_s^2$. However,  this effect shown in Fig.\,\ref{fig:fac} is numerically small(about 3\%$\approx 1/(10N_c))$ of the sum of factorized $\oplus$ non-factorized contributions as expected from large $N_c$ limit and Fierz transformations. This feature has been already observed explicitly in our previous work\,\cite{SU3,MOLE12}. This small effect of the non-factorized contribution justifies the accuracy of our approximation by only using the factorized diagrams in the NLO perturbative contributions.

   We do not also see the relevance / appearance of the Landau singularities mentioned by\,\cite{LUCHA1,LUCHA2} in the analysis using the OPE in the Euclidian region. However, the two-point function analyzed in\,\cite{LUCHA1,LUCHA2} has nothing to do with the one analyzed in our paper as it corresponds to a four-point function compacted into a two-point function but with four legs i.e with two incoming and two-outgoing momenta. This four-point function is more relevant for the analysis of hadron-hadron scatterings (see the example of $\pi\pi$ and $\gamma\gamma$ in\,\cite{MENES1,MENES2}), while in this case, a two-point function enters differently via a gluonium intermediate state\,\cite{MENES3}.

   From the analysis of Eq.\,\ref{eq:cont}, we have shown that the postulated lowest mass ground state dominates the spectral function. This feature indicates that the non-resonant states do not play a crucial role in the analysis. This conclusion may go in line with the answer of\,\cite{WANG1,WANG2} on some of the comments of\,\cite{LUCHA1,LUCHA2}. 
\begin{table*}[hbt]
\setlength{\tabcolsep}{0.1pc}
\catcode`?=\active \def?{\kern\digitwidth}
 {\footnotesize
\begin{tabular*}{\textwidth}{@{}l@{\extracolsep{\fill}}|cccccc cc cc} 
\hline
\hline
      Scalar          & \multicolumn{5}{c}{$M_{\bar c\bar ccc}$ [GeV]} 
                  & \multicolumn{5}{c}{$M_{\bar b\bar bbb}$ [GeV]} \\
\cline{2-6} \cline{7-11}
   $ \bar q\bar q qq$           & \multicolumn{1}{c}{LO} 
       & \multicolumn{1}{c}{NLO} 
         & \multicolumn{1}{c}{NLO $\oplus$ G3} 
                 & \multicolumn{1}{c}{LO\,\cite{STEELE}} 
                 & \multicolumn{1}{c}{ LO\,\cite{WANG} } 
                & \multicolumn{1}{c}{LO} 
       & \multicolumn{1}{c}{NLO}    
         & \multicolumn{1}{c}{NLO $\oplus$ G3}               
       & \multicolumn{1}{c}{LO\,\cite{STEELE}} 
                 & \multicolumn{1}{c}{LO\,\cite{WANG} }                
    
                       \\
\cline{1-6} \cline{7-11} 
\bf Eq.\,\ref{eq:tetra-steele}\\
$\overline S_qS_q$&6.59&$6.39\pm 0.08$&$6.41\pm 0.08$&$6.44\pm 0.15$&  &19.51&$19.13\pm 0.08$&$19.22\pm 0.12$&$18.45\pm0.15$&  \\ 
$\bar A_qA_q$ &6.52&$6.49\pm 0.07$&$6.45\pm 0.08$&$6.46\pm 0.16$&$-$&19.51&$19.93\pm 0.15$&$19.87\pm 0.16$&$18.46\pm 0.14$&$-$\\ 
$\overline V_qV_q$ &6.55&$6.61\pm 0.09$&$6.46\pm0.18$&$6.59\pm 0.17$&  &19.49&$19.53\pm 0.07$&$19.49\pm 0.08$&$18.59\pm 0.17$&  \\
$\overline P_qP_q$&7.37&$7.05\pm 0.07$&$6.80\pm 0.27$&$6.82\pm 0.18$&  &19.96&$19.78\pm 008$&$19.75\pm 0.08$&$19.64\pm 0.14$&  \\ 
\cline{0-0}
\bf Eq.\,\ref{eq:tetra-wang}\\
$\bar A_qA_q$ &6.50&$6.51\pm 0.06$&$6.47\pm 0.07$&  &$5.99\pm 0.08$& 19.49&$19.75\pm 0.11$&$19.72\pm 0.12$&  &$18.84\pm0.09$ \\
\hline\hline
\end{tabular*}
}
 \caption{Comparison of the values of the $0^{++}$ scalar tetraquark masses  and couplings from different QSSR approaches. Our predictions are at LO (only the central value is quoted) and up to NLO of PT series where the errors come from\, Table\,\ref{tab:res}. The predictions of Ref.\,\cite{STEELE} are from Moments at LO and of Ref.\,\cite{WANG} from LSR at LO. As already mentioned earlier, we notice that the choice of the numerical values of the $\overline{MS}$ running quark masses used at LO is not justified due to the ambiguous quark mass definition to that order. One may also have equally used a pole / on-shell mass which naturally appears in the expression of the spectral function evaluated using on-shell quark mass. }  
\label{tab:comparison}
\end{table*}

\subsection{Systematic errors}
   As mentioned in Section\,\ref{sec:spectral}, one expects that at the optimization region, an eventual  duality violation is expected to be negligible and the QCD continuum contribution which parametrizes non-resonant states is dominated by the lowest resonance as can be checked  from Eq.\,\ref{eq:cont}. Therefore, the high-energy tail of the spectral function cannot bring a sizeable systematic error.

    The error due to the truncation of the PT series cannot be quantified with a good accuracy as the LO contributions are quite sensitive to the quark mass definition (pole or running) in some other channels. Using an approach similar to the one leading to Eq.\,\ref{eq:pt-xc} where a geometric growth of the $a_s$-coefficients has been assumed, we deduce the error estimate in Table\,\ref{tab:res}.

   We have estimated  the unknown higher dimension condensates contributions in the OPE quoted in Table\,\ref{tab:res} as discussed in Section\,\ref{sec:chi0}. 
\subsection*{\b New compared with available QSSR results}

Compared to previous QSSR LO results given in the literature (see Table\,\ref{tab:comparison}):

   We have included (for the first time) the NLO corrections which is mandatory for giving a sense on the definition and numerical values of the input heavy quark mass which plays a crucial role in the analysis. 

   We have added the contributions of the dimension-six $\la G^3\ra$ condensates, which are quite large
for the ${\eta_{q}\eta_{q}}$ and ${J/\psi J/\psi,\Upsilon\Upsilon}$ molecules  and  for the $\bar V_qV_q$ and $\bar P_qP_q$ tetraquark states.

   Our results are shown in Table\,\ref{tab:res} where systematic analysis of some possible configurations of the $0^{++}$ molecule and four-quark states have been done. 

\subsection{LSR versus the ratio of MOM results}

   Taking, the example of the ${\overline\chi_{c0}}\chi_{c0}$ molecule and $\bar S_qS_q$-tetraquark, we use the ratio of moments as in\,\cite{STEELE}:
\bea
M^2_{\cal T,M}&=&\frac{{\cal M}_n(Q^2_0)}{{\cal M}_{n+1}(Q^2_0)}-Q^2_0\,\, : \nnb\\
{\cal M}_n(Q^2_0)&=&\frac{1}{\pi}\int_{16m_Q^2}^{\infty}\hspace*{-0.25cm}dt\,\frac{{\rm Im}\,\Pi_{\cal T,M}(t)}{(t+Q^2_0)^n}~,
\eea
where $M_{\cal T,M}$ is the molecule or tetraquark mass. We take e.g $Q^2_0=4m_Q^2$. 

Then,  we find that the 
LO and LO $\oplus$ NLO results are about the same as from the LSR obtained in the previous sections. To NLO and including $\la\alpha_sG^2\ra$, one obtains in units of GeV\,: 
\beq
M_{\chi_{c0}\chi_{c0}}\simeq 6.93,~~M_{S_cS_c}\simeq 6.38, ~~M_{S_bS_b}\simeq 19.29~,
\eeq
compared to the ones from LSR in Table\,\ref{tab:res}, indicating that the two methods give (within the errors) the same results.
\subsection{On the ratio of MOM results of Ref.\,\cite{STEELE}}
   Using the QCD expression of the $\bar S_qS_q$ tetraquark two-point function given in Appendix A, we have also compared our LO $\oplus$ $\la \alpha_s G^2\ra$  MOM results :
\beq
M_{S_cS_c}\simeq 6.78~{\rm GeV}, ~~M_{S_bS_b}\simeq 19.53~~{\rm GeV},
\label{eq:ss}
\eeq
with our LO LSR results given in Table\,\ref{tab:comparison} where we find (within the errors) a good agreement. 

   However, by comparing these LO MOM results with the ones from\,\cite{STEELE} quoted in  Table\,\ref{tab:comparison}, one can see
that the results of\,\cite{STEELE} are about 0.34 GeV (resp. 1.08 GeV) for the charm (resp. beauty) case lower than the ones in Eq.\,\ref{eq:ss}. More generally, compared to our LO ones, the LO results of\,\cite{STEELE}  have the tendancy to underestimate the mass results. 

   With the inclusion of the NLO QCD corrections, our predictions  agree (within the errors) with the LO results of\,\cite{STEELE} for the charm and for the $\bar P_qP_q$ beauty channels.  For the $\bar S_qS_q,~\bar A_qA_q$ and $\bar V_qV_q$ beauty ones, the  disagreements persist and range from 0.77 to 1.41 GeV. We cannot trace back the origin of a such discrepancy as the comparison of the QCD expressions of the full correlator given in\,\cite{STEELE} with the one using the spectral function is not easy due to the choice of variables used by the authors. 

   Therefore, unlike Ref.\,\cite{STEELE}, we expect, like in the charm case, that the tetraquark beauty states are above the $\eta_b\eta_b$ and $\Upsilon(1S)\Upsilon(1S)$ thresholds. The future experimental findings of these beauty states may select among these theoretical predictions. 
\subsection{Comparison with the LSR results of Ref.\,\cite{WANG}}
   We have also compared our results for the $\bar A_qA_q$ scalar tetraquark with the LO ones of\,\cite{WANG} using the current in Eq.\,\ref{eq:tetra-wang}. 

   The PT QCD expressions agree each other at LO. There is a slight difference for the  $\la \alpha_s G^2\ra$ contribution for higher values of $t$ to all values of the heavy quark mass but this difference  affects only slightly the predictions.  

   At LO and including the $\la \alpha_sG^2\ra$ contribution, our values of the $\bar A_qA_q$  couplings of about 287 (resp. 78)  keV for the charm (resp. bottom) are comparable with the ones of\,\cite{WANG} (289 (resp. 54) keV) if the (unjustified) choice of $\overline{MS}$-mass is used. 

   For the charm, the $\bar A_qA_q$ mass of\,\cite{WANG} is (460  550) MeV lower than the one of\,\cite{STEELE} and our LO result, while for the bottom it is 670 MeV lower than our LO result but 380 MeV higher than that of\,\cite{STEELE}\,(see Table\,\ref{tab:comparison}). However, the origin of  this discrepancy  does not come from  the QCD input parameters as we use about the same values. This example puts a question mark on the unusual treatment of the sum rules by the author in\,\cite{WANG}. 

   His choice of the subtraction scale $\mu\simeq (1.2\sim 2.2)$ GeV for the charm (resp. $(2.3\sim 3.3)$ GeV for the bottom)  based, for instance, on the identification of the sum of the PT running mass $(\overline{m}_c+\overline{m}_b)(\mu)$ with the value of the $B_c$-mass\,\cite{WANG3} is difficult to justify in the absence of NP-contributions (binding energy).  However, such low values of $\mu$ are quite dangerous as, at this low scale, the PT radiative corrections are expected to be large and can strongly affect the final result. This is indeed the case for the coupling where, at the $\mu$-stability (4.5 GeV for the charm  and 7.25 GeV for the bottom) the NLO corrections increase it by 59\% for the charm and 83\% for the bottom. This effect is obviously larger for smaller values of $\mu$. 

   Moreover, using only the $\mu$-dependence of the running values of $\alpha_s$ and $m_Q$ into the PT LO expression of the sum rule is also inconsistent while the identification of the QCD continuum threshold with the mass of the first radial excitation can be inaccurate as the QCD continuum is expected to smear all higher state contributions.

   It is also remarkable to notice from Tables\,\ref{tab:res} and\,\ref{tab:comparison} the (almost) independence of our results on the form of the current for the $\bar A_qA_q$ tetraquark. 

   For a consistency check of our results, we compare our result for the $\bar A_qA_q$ tetraquark mass $M_{\bar A_qA_q}\simeq 6.47$ (resp. 19.72) GeV from the current of\,\cite{WANG} within a $\bar 3_c \otimes 3_c$ color representation with the one from the combination of molecule currents 
$2(\bar S_qS_q+\bar P_qP_q)+\bar V_qV_q-\bar A_qA_q$ given there. Using a quadratic mass relation, we deduce  at NLO $\oplus G3$: $M_{\bar A_qA_q}\simeq 6.38$ (resp. 19.49) GeV in agreement (within the errors) with our predicted tetraquark masses. 
\subsection{Some phenomenological implications}
One can notice from Tables\,\ref{tab:res} and\,\ref{tab:comparison} that :

   Our different QSSR predictions cannot disentangle (within the errors) the mass of a molecule from a tetraquark state as already found in some of our previous works\,\cite{HEP18,SU3,QCD16,MOLE16,X5568}. 

   Our results do not favour the ones from some potential models where the exotic states are below the $\overline{\eta}_c\eta_c$ meson thresholds. Instead, our results
may explain the existence of a  $0^{++}$ broad structure around (6.2    6.7) GeV  which can be due to $\overline{\eta}_c{\eta}_{c}$, $\overline{\chi}_{c1}{\chi}_{c1}$  and $ \,\overline{J/\psi}{J/\psi}$ molecules or /and to scalar-scalar, vector-vector and axial-axial scalar tetraquark states.  

   If the new LHCb peak candidate\,\cite{LHCb1,LHCb2} around (6.8    6.9) GeV is a $0^{++}$ state, the value of its mass suggests that it is likely a $\overline{\chi}_{c0}{\chi}_{c0}$ molecule or a pseudoscalar-pseudoscalar tetraquark states. Its signature from a $\overline{J/\psi}{J/\psi}$ invariant mass may come  from the di-${\chi}_{c0}$ decaying to di-$\gamma J/\psi$. 

   In the case of a $\overline{\chi}_{c1}{\chi}_{c1}$ molecule, the predicted mass is below the ${\chi}_{c1}{\chi}_{c1}$ threshold while our NLO predictions for the beauty states indicate that all of them are above the $\eta_b\eta_b$ and $\Upsilon(1S)\Upsilon(1S)$ thresholds. 

We plan to calculate the spectra of some other $0^{-}$, $1^{\pm}$ and $2^{++}$ channels  and eventually their widths in a future work. 
\vspace*{-0.5cm}
\appendix
\section{$\la G^3\ra$ contributions to the $\overline\chi_{0q}\chi_{0q}$ spectral function}
\begin{figure}[hbt]
\begin{center}
\hspace*{0.cm}\includegraphics[width=8cm]{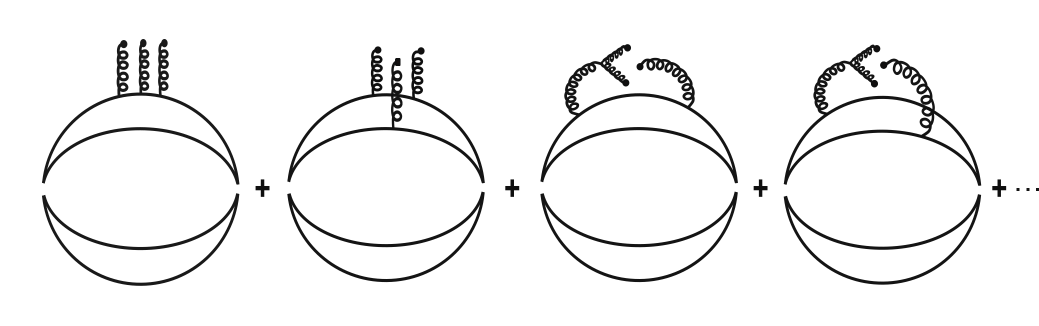}
\vspace*{-0.25cm}
\caption{\footnotesize  Factorised $\la G^3\ra$ contribution to the spectral function.} 
\label{fig:fg3}
\end{center}
\vspace*{-0.5cm}
\end{figure}

\begin{figure}[hbt]
\begin{center}
\hspace*{0.cm}\includegraphics[width=8cm]{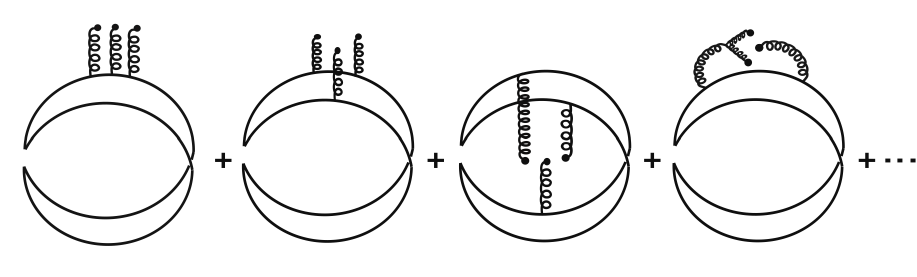}
\vspace*{-0.25cm}
\caption{\footnotesize  Non-factorised $\la G^3\ra$ contribution to the spectral function.} 
\label{fig:g3nf}
\end{center}
\end{figure}

The $\la G^3\ra$ contributions to the $\bar\chi_{0q}\chi_{0q}$ spectral function are given by the Feynman diagrams drawn in Figs.\,\ref{fig:fg3} and\,\ref{fig:g3nf}. As the expression is quite lengthy, we shall only present the one for $\overline\chi_{0q}\chi_{0q}$ but not for some other molecules. 
For convenience, the spectral function is parametrized (here and in the following)  in terms of the variables $x,y,z$ and the corresponding limits of integration defined in Eq.\,\ref{eq:var}.  The parameter $\epsilon$ is equal to zero for factorised and to one for the total (factorised $\oplus$ non-factorised) contributions. The $\la G^3\ra$ contributions read: 
\begin{widetext}
\bea
&&\frac{1}{\pi}\mbox{Im}\Pi^{S\,;\,\GGG}_{\overline{\chi}_{q0}\chi_{q0}}(t)=\frac{\gGGG}{3^2 \times 2^9 \pi^6}\int_{x\,y\,z}\frac{1}{y^3 z^2} \Big(480 M^2 x\,z^2 y^4-1200 t\,x\,z^2 y^4+108 m^2 x\,y^4-60 M^2 x\,z\,y^4+120 t\,x\,z\,y^4\nnb \\
&&+480 M^2 x\, z^3 y^3-1200 t\,x\,z^3 y^3\hspace*{0.08cm}+\hspace*{0.08cm}108 m^2 x^2 y^3\hspace*{0.08cm}-\hspace*{0.08cm}120 m^2 z^2 y^3+480 M^2 x^2 z^2 y^3-1200 t\, x^2 z^2 y^3-600 M^2 x\,z^2 y^3 \nnb \\
&&+1440 t\,x\,z^2 y^3-20 m^2 x\,y^3+ \hspace*{0.1cm} 20 m^2 z\,y^3-60 M^2 x^2 z\, y^3 + 120 t\,x^2 z\,y^3+270 m^2 x\,z\,y^3+30 M^2 x\,z\,y^3-75 t\,x\,z\,y^3 \nnb \\
&& -60 M^2 x\,z^3 y^2\hspace*{-0.1cm}+120 t\,x\,z^3 y^2\hspace*{-0.1cm}+88 m^2 x^2 y^2 \hspace*{-0.1cm}+20 m^2 z^2 y^2-60 M^2 x^2 z^2 y^2 \hspace*{-0.1cm}+120 t\,x^2 z^2 y^2 \hspace*{-0.1cm} +162 m^2 x\,z^2 y^2 \hspace*{-0.1cm}+30 M^2 x\,z^2 y^2 \nnb \\
&&-75 t\,x\,z^2 y^2-88 m^2 x\,y^2+162 m^2 x^2 z\,y^2-30 M^2 x^2 z\,y^2+45 t\,x^2 z\,y^2 - \hspace*{0.05cm}94 m^2 x\,z\, y^2\hspace*{0.05cm}+\hspace*{0.05cm}30 M^2 x\,z\, y^2\hspace*{0.05cm}-\hspace*{0.05cm}45 t\,x\,z\, y^2\hspace*{-0.05cm} \nnb \\
&&+108 m^2 x\, z^3 y\hspace*{-0.05cm}-\hspace*{-0.05cm}20 m^2 x\,z^2 y\hspace*{-0.05cm}-\hspace*{-0.05cm}20 m^2 x^2 z\, y+20 m^2 x\, z\, y\hspace*{-0.08cm}+\hspace*{-0.08cm}108 m^2 x\, z^4\hspace*{-0.08cm}+\hspace*{-0.08cm}108 m^2 x^2 z^3\hspace*{-0.08cm}-\hspace*{-0.08cm}216 m^2 x\, z^3\Big)\nnb\\
&&+\frac{\gGGG }{23040\, \pi^6}\int_{x\,y\,z}\hspace*{-0.18cm}\frac{e^{-M^2\tau}}{y^5 z^3} \hspace*{0.05cm} \Big(54 z^3 \tau^2 m^8-54 M^2 x\, z^4 \tau^2 m^6-54 M^2 y\, z^4 \tau^2 m^6\hspace*{-0.05cm}-\hspace*{-0.08cm}180 y^4 \tau\, m^6\hspace*{-0.05cm}-\hspace*{-0.05cm}315 y\, z^3 \tau\,  m^6\hspace*{-0.05cm}+\hspace*{-0.05cm}25 y^3 z\, \tau\,  m^6 \nnb \\
&&-180 x\, y^5 m^4\hspace*{-0.05cm}-\hspace*{-0.05cm}108 x\,y\,z^5 m^4 \hspace*{-0.05cm} - \hspace*{-0.05cm}180 x^2 y^4 m^4 \hspace*{-0.05cm} + \hspace*{-0.05cm} 180 x\, y^4 m^4\hspace*{-0.05cm} - \hspace*{-0.05cm}108 x\, y^2 z^4 m^4 \hspace*{-0.05cm}+ \hspace*{-0.05cm}90 y^2 z^4 m^4 \hspace*{-0.05cm} - \hspace*{-0.05cm}108 x^2 y\, z^4 m^4+423 x\, y\, z^4 m^4 \nnb \\
&&+270 y^2 z^3 m^4 \hspace*{-0.05cm} + \hspace*{-0.05cm} 135 y^4 z^2 m^4  +  25 x\, y^3 z^2 m^4 \hspace*{-0.05cm} - \hspace*{-0.05cm} 50 y^3 z^2 m^4 \hspace*{-0.05cm} - \hspace*{-0.05cm}54 M^4 x\, y\, z^5 \tau^2 m^4 \hspace*{-0.05cm} - \hspace*{-0.05cm} 54 M^4 x\, y^2 z^4 \tau^2 m^4 \hspace*{-0.05cm} - \hspace*{-0.05cm} 54 M^4 x^2 y\, z^4 \tau^2 m^4\nnb \\
&&+54 \, M^4 x\, y\, z^4 \tau^2 m^4 \hspace*{0.05cm} + \hspace*{0.05cm} 90 y^5 z\, m^4-155 x\, y^4 z\, m^4 \hspace*{0.05cm}+ \hspace*{0.05cm} 220 y^4 z\, m^4+25 x^2 y^3 z\, m^4 \hspace*{0.05cm} - \hspace*{0.05cm} 25 x\, y^3 z\, m^4 \hspace*{0.05cm} - \hspace*{0.05cm}180 \, M^2 x\, y^5 \tau\, m^4 \nnb \\
&&-108 M^2 x\, y\, z^5 \tau\, m^4\hspace*{-0.15cm}-\hspace*{-0.09cm}180 M^2 x^2 y^4 \tau\,  m^4\hspace*{-0.15cm}+\hspace*{-0.09cm}180 M^2 x\, y^4 \tau\,  m^4\hspace*{-0.15cm}+\hspace*{-0.08cm}90 M^2 y^2 z^4 \tau\,  m^4\hspace*{-0.15cm}-\hspace*{-0.09cm}108 M^2 x\, y^2 z^4 \tau\,  m^4-108 M^2 x^2 y\, z^4 \tau\, m^4\hspace*{-0.08cm} \nnb \\
&&+423 M^2 x\, y\, z^4 \tau\,  m^4+135 M^2 y^4 z^2 \tau\,  m^4\hspace*{-0.08cm}+25 M^2 x\, y^3 z^2 \tau\,  m^4\hspace*{-0.08cm}+\hspace*{-0.08cm}90 M^2 y^5 z\, \tau\,  m^4\hspace*{-0.08cm}-\hspace*{-0.08cm}155 M^2 x\, y^4 z \tau\,  m^4\hspace*{-0.08cm}+\hspace*{-0.08cm}25 M^2 x^2 y^3 z\, \tau\,  m^4 \nnb \\
&&-25 M^2 x\, y^3 z\, \tau\,  m^4 \hspace*{-0.05cm} + \hspace*{-0.05cm} 360 M^2 x\, y^2 z^5 m^2 \hspace*{-0.05cm}+ \hspace*{-0.05cm} 360 M^2 x\, y^3 z^4 m^2 \hspace*{-0.05cm} +  360 M^2 x^2 y^2 z^4 m^2 \hspace*{-0.05cm} -  630 M^2 x\, y^2 z^4 m^2-400 M^2 y^5 z^3 m^2 \nnb \\
&&+50 \, M^2 y^4 z^3 m^2+540 \,  M^2 x\, y^4 z^3 m^2-50\, M^2 x\, y^3 z^3 m^2+50 \, M^2 y^5 z^2 m^2+900 \, M^2 x\,\, y^5 z^2 m^2+540 \, M^2 x^2 y^4 z^2 m^2 \nnb \\
&&-370 M^2 x\, y^4 z^2 m^2-50 M^2 x^2 y^3 z^2 m^2+50 M^2 x\, y^3 z^2 m^2+360 M^2 x\, y^6 z\, m^2+360 M^2 x^2 y^5 z\, m^2-\hspace*{-0.07cm}140 M^2 x\, y^5 z\, m^2 \nnb \\
&&+220 M^2 x^2 y^4 z\, m^2 \hspace*{-0.05cm} - \hspace*{-0.05cm}220 M^2 x\, y^4 z\, m^2 \hspace*{-0.05cm} + \hspace*{-0.05cm} 90 M^4 x\, y^2 z^5 \tau\,m^2 \hspace*{-0.05cm}+ \hspace*{-0.05cm}90 M^4 x\, y^3 z^4 \tau\,  m^2\hspace*{-0.12cm}+\hspace*{-0.08cm}90 M^4 x^2 y^2 z^4 \tau\,  m^2\hspace*{-0.12cm}-\hspace*{-0.08cm}90 M^4 x\, y^2 z^4 \tau\,  m^2\hspace*{-0.12cm} \nnb \\
&&-100 \, M^4 \, y^5 \, z^3 \,\tau\, m^2 \hspace*{0.05cm}+\hspace*{0.05cm}135 \, M^4 x\, y^4 z^3 \tau\,  m^2 \hspace*{0.05cm}+ \hspace*{0.05cm} 225 \, M^4 x\, y^5 z^2 \tau\,  m^2 \hspace*{0.05cm}+ \hspace*{0.05cm}135 \, M^4 x^2 y^4 z^2 \tau\,  m^2 \hspace*{0.05cm}- \hspace*{0.05cm} 135\, M^4 x\,\, y^4 \, z^2 \, \tau\,  m^2 \nnb \\
&&+\hspace*{-0.08cm}90 M^4 x\, y^6 z\, \tau\,  m^2\hspace*{-0.12cm}+\hspace*{-0.08cm}90 M^4 x^2 y^5 z\, \tau\,  m^2\hspace*{-0.12cm}-\hspace*{-0.08cm}90 M^4 x\, y^5 z\, \tau\,  m^2\hspace*{-0.12cm}-\hspace*{-0.08cm}900 M^4 x\, y^5 z^4\hspace*{-0.12cm}+\hspace*{-0.12cm}50 M^4 x\, y^4 z^4\hspace*{-0.12cm}-\hspace*{-0.08cm}900 M^4 x\, y^6 z^3\hspace*{-0.12cm}-\hspace*{-0.08cm}900 M^4 x^2 y^5 z^3 \nnb \\
&&+1000 M^4 x\, y^5 z^3 +50 M^4 x^2 y^4 z^3-50 M^4 x\, y^4 z^3+50 M^4 x\, y^6 z^2+50 M^4 x^2 y^5 z^2-50 M^4 x\, y^5 z^2-100 M^6 x\, y^5 z^4 \tau \nnb \\
&&-100 M^6 x\, y^6 z^3 \tau-100 M^6 x^2 y^5 z^3 \tau +100 M^6 x\, y^5 z^3 \tau \Big)\nnb
\eea
\bea
&&-\frac{\epsilon\, \gGGG}{3^2\times 2^{13} \pi ^6}\int_{x\,y\,z} \frac{1}{y^3 z^2} \Big(2080 M^2 x\, z^2 y^4 \hspace*{-0.05cm}-\hspace*{-0.05cm}5200 t\, x\, z^2 y^4 \hspace*{-0.05cm}+ \hspace*{-0.05cm} 472 m^2 x\, y^4 \hspace*{-0.05cm} - \hspace*{-0.05cm} 180 M^2 x\, z\, y^4 \hspace*{-0.05cm} + \hspace*{-0.05cm}360 t\, x\, z\, y^4 \hspace*{-0.05cm} + \hspace*{-0.05cm} 2080 M^2 x\, z^3 y^3 \nnb \\
&&-5200 t\, x\, z^3 y^3+472 m^2 x^2 y^3-\hspace*{0.05cm}480 m^2 z^2 y^3+\hspace*{0.05cm}2080 M^2 x^2 z^2 y^3-\hspace*{0.05cm}5200 t\, x^2 z^2 y^3-\hspace*{0.05cm}2440 M^2 x\, z^2 y^3\hspace*{0.05cm}+\hspace*{0.05cm}5920 t\, x\, z^2 y^3\hspace*{0.05cm} \nnb \\
&&-408 m^2 x\, y^3 \hspace*{-0.05cm}+ 32 m^2 z\, y^3 \hspace*{-0.05cm}-  180 M^2 x^2 z\, y^3 \hspace*{-0.05cm}+360 t\, x^2 z\, y^3 \hspace*{-0.05cm} +  904 m^2 x\, z\, y^3 \hspace*{-0.05cm} +  90 M^2 x\, z\, y^3 \hspace*{-0.05cm} - 225 t\, x\, z\, y^3-180 M^2 x\, z^3 y^2 \nnb \\
&&+360 t\, x\, z^3 y^2 \hspace*{-0.05cm} + \hspace*{-0.05cm} 208 m^2 x^2 y^2 \hspace*{-0.05cm} + \hspace*{-0.05cm} 80 m^2 z^2 y^2 \hspace*{-0.05cm} -180 M^2 x^2 z^2 y^2 \hspace*{-0.05cm} +360 t\, x^2 z^2 y^2 \hspace*{-0.05cm} +184 m^2 x\, z^2 y^2 \hspace*{-0.05cm} + \hspace*{-0.05cm} 90 M^2 x\, z^2 y^2 \hspace*{-0.05cm} - \hspace*{-0.05cm}225 t\, x\, z^2 y^2 \nnb \\
&&-208 m^2 x\, y^2 \hspace*{-0.05cm} +432 m^2 x^2 z\, y^2 \hspace*{-0.05cm} -90 M^2 x^2 z\, y^2 \hspace*{-0.05cm}+ \hspace*{-0.05cm} 135 t\, x^2 z\, y^2-332 m^2 x\, z\, y^2+90 M^2 x\, z\, y^2\hspace*{-0.08cm}-\hspace*{-0.08cm}135 t\, x\, z\, y^2\hspace*{-0.08cm}+\hspace*{-0.08cm}184 m^2 x\, z^3 y\hspace*{-0.08cm} \nnb \\
&&+40 m^2 x^2 z^2 y\hspace*{-0.08cm}-\hspace*{-0.08cm}100 m^2 x\, z^2 y\hspace*{-0.08cm}-\hspace*{-0.08cm}60 m^2 x^2 z\, y\hspace*{-0.08cm}+\hspace*{-0.08cm}60 m^2 x\, z\, y\hspace*{-0.08cm}+\hspace*{-0.08cm}144 m^2 x\, z^4+144 m^2 x^2 z^3-288 m^2 x\, z^3\Big)\nnb \\
&&-\frac{\epsilon\, \gGGG}{552960\, \pi^6} \hspace*{0.1cm} \int_{x\,y\,z} \frac{e^{-M^2\tau}}{y^5 z^3} \hspace*{0.05cm} \Big(108 \, z^3 \tau^2 m^8 \hspace*{0.05cm} - \hspace*{0.05cm} 108 \, M^2 x\, z^4 \tau^2 m^6 \hspace*{0.05cm} - \hspace*{0.05cm} 108 \, M^2 y\, z^4 \tau^2 m^6 \hspace*{0.05cm} - \hspace*{0.05cm}1080 \, y^4 \tau\,  m^6-630 \, y\, z^3 \tau\,  m^6 \nnb \\
&&+60 y^3 z\, \tau\,  m^6 \hspace*{-0.05cm} - \hspace*{-0.05cm} 360 x\, y^5 m^4 \hspace*{-0.05cm} - \hspace*{-0.05cm} 216 x\, y\, z^5 m^4 \hspace*{-0.05cm} -720 x^2 y^4 m^4 \hspace*{-0.05cm} +720 x\, y^4 m^4-216 x\, y^2 z^4 m^4 \hspace*{-0.05cm} +180 y^2 z^4 m^4 \hspace*{-0.05cm} -216 x^2 y\, z^4 m^4 \nnb \\
&&+846 \, x\, y\, z^4 m^4 \hspace*{-0.05cm} + \hspace*{-0.05cm} 540 \, y^2 z^3 m^4+810 \, y^4 z^2 m^4+190 \, x\, y^3 z^2 m^4\hspace*{-0.05cm}-\hspace*{-0.05cm}120 y^3 z^2 m^4\hspace*{-0.05cm}-\hspace*{-0.05cm}108 M^4 x\, y\, z^5 \tau ^2 m^4\hspace*{-0.12cm}-\hspace*{-0.08cm}108 M^4 x\, y^2 z^4 \tau^2 m^4\hspace*{-0.12cm} \nnb \\ 
&&-108 M^4 x^2 y\, z^4 \tau^2 m^4 \hspace*{-0.05cm}+ \hspace*{-0.05cm} 108 M^4 x\, y\, z^4 \tau^2 m^4+180 y^5 z\, m^4-170 x\, y^4 z\, m^4+1500 y^4 z\, m^4\hspace*{-0.1cm}+\hspace*{-0.08cm}100 x^2 y^3 z\, m^4\hspace*{-0.1cm}-\hspace*{-0.08cm}100 x\, y^3 z\, m^4\hspace*{-0.1cm} \nnb \\
&&-360 M^2 x\, y^5 \tau\,  m^4-\hspace*{-0.08cm}216 M^2 x\, y\, z^5 \tau\,m^4-\hspace*{-0.08cm}720 M^2 x^2 y^4 \tau\, m^4\hspace*{-0.1cm}+\hspace*{-0.08cm}720 M^2 x\, y^4 \tau\, m^4\hspace*{-0.1cm}+\hspace*{-0.08cm}180 M^2 y^2 z^4 \tau\, m^4 \hspace*{-0.05cm} -  216 M^2 x\, y^2 z^4 \tau\, m^4\hspace*{-0.1cm} \nnb \\
&&-216 M^2 x^2 y\, z^4 \tau\, m^4\hspace*{-0.1cm}+ \hspace*{-0.08cm} 846 M^2 x\, y\, z^4 \tau\,  m^4\hspace*{-0.1cm}+\hspace*{-0.08cm}810 M^2 y^4 z^2 \tau\, m^4\hspace*{-0.1cm}+\hspace*{-0.08cm}190 M^2 x\, y^3 z^2 \tau\,  m^4\hspace*{-0.1cm}+\hspace*{-0.08cm}180 M^2 y^5 z\, \tau\,  m^4\hspace*{-0.1cm}-\hspace*{-0.08cm}170 M^2 x\, y^4 z\, \tau\,  m^4\nnb\\
&&+100 M^2 x^2 y^3 z\, \tau\,  m^4\hspace*{-0.1cm}-\hspace*{-0.08cm}100 M^2 x\, y^3 z\, \tau\,  m^4\hspace*{-0.1cm}+\hspace*{-0.08cm}720 M^2 x\, y^2 z^5 m^2\hspace*{-0.1cm}+\hspace*{-0.08cm}920 M^2 x\, y^3 z^4 m^2\hspace*{-0.1cm}+\hspace*{-0.08cm}720 M^2 x^2 y^2 z^4 m^2\hspace*{-0.08cm}-\hspace*{-0.08cm}1260 M^2 x\, y^2 z^4 m^2 \nnb \\
&&-2400 M^2 y^5 z^3 m^2+300 M^2 y^4 z^3 m^2+  920 M^2 x\, y^4 z^3 m^2+ \hspace*{0.05cm}200 M^2 x^2 y^3 z^3 m^2- \hspace*{0.05cm} 425 M^2 x\, y^3 z^3 m^2+ \hspace*{0.05cm} 120 M^2 y^5 z^2 m^2 \nnb \\
&&+ 4520 \hspace*{0.05cm} M^2 \,x\, \, y^5 \, z^2 \, m^2\hspace*{0.08cm} + \hspace*{0.04cm}2160 \hspace*{0.05cm} M^2 \, x^2 \, y^4 \, z^2 \, m^2 \hspace*{0.08cm}- \hspace*{0.08cm} 1785 \hspace*{0.05cm}  M^2 \, x\, y^4 \, z^2 m^2\hspace*{0.08cm}-\hspace*{0.08cm}225 \hspace*{0.05cm}  M^2 x^2 y^3 z^2 m^2\hspace*{0.08cm}+\hspace*{0.08cm}225 \hspace*{0.05cm} M^2 \, x\, y^3 \, z^2 \, m^2\hspace*{0.08cm} \nnb \\
&&+2360 M^2 \hspace*{0.05cm} x\, y^6 \, z\, m^2 \hspace*{0.12cm} + \hspace*{0.1cm}2360 \hspace*{0.05cm}  M^2 \, x^2 \,  y^5 \, z\, m^2\hspace*{0.12cm}- \hspace*{0.1cm} 2120 \, M^2\hspace*{0.05cm} x\, y^5 z\, m^2\hspace*{0.12cm}+\hspace*{0.1cm}780 \hspace*{0.05cm} M^2 \, x^2 \, y^4 \, z\, m^2\hspace*{0.12cm}-\hspace*{0.1cm}780 \hspace*{0.05cm} M^2 \, x\, y^4 \, z\, m^2 \nnb \\
&&+180 \hspace*{0.05cm} M^4 x\, y^2 z^5 \tau\, m^2\hspace*{0.12cm}+\hspace*{0.08cm}230 \hspace*{0.05cm} M^4 x\, y^3 z^4 \tau\, m^2\hspace*{0.12cm}+\hspace*{0.1cm}180 \hspace*{0.05cm} M^4 x^2 y^2 z^4 \tau\, m^2\hspace*{0.12cm}-\hspace*{0.1cm}180 \hspace*{0.05cm} M^4 x\, y^2 z^4 \tau\,  m^2\hspace*{0.12cm}-\hspace*{0.08cm}600 \hspace*{0.05cm} M^4 y^5 z^3 \tau\,  m^2\nnb \\
&&+230 M^4 x\, y^4 z^3 \tau\, m^2+590 M^4 x^2 y^5 z\, \tau\,  m^2-590 M^4 x\, y^5 z\, \tau\,  m^2-5850 M^4 x\, y^5 z^4+225 M^4 x\, y^4 z^4-5850 M^4 x\, y^6 z^3 \nnb \\
&&-5850 M^4 x^2 y^5 z^3 \hspace*{-0.05cm} + \hspace*{-0.05cm} 6300 M^4 x\, y^5 z^3 \hspace*{-0.05cm} + \hspace*{-0.05cm} 225 M^4 x^2 y^4 z^3 \hspace*{-0.05cm} + 50\, M^4 x^2 y^3 z^3 \tau\,m^2 \hspace*{-0.05cm} -  50\, M^4 x\, y^3 z^3 \tau\,m^2 \hspace*{-0.05cm} + 1130\, M^4 x\, y^5 z^2 \tau\,  m^2 \nnb \\
&&+540\, M^4 x^2 y^4 z^2 \tau\,  m^2 \hspace*{-0.05cm} -540\, M^4 x\, y^4 z^2 \tau\,m^2+590\, M^4 x\, y^6 z\, \tau\,  m^2-225 M^4 x\, y^4 z^3+225 M^4 x\, y^6 z^2+225 M^4 x^2 y^5 z^2 \nnb \\
&& -225 M^4 x\, y^5 z^2-650 M^6 x\, y^5 z^4 \tau -650 M^6 x\, y^6 z^3 \tau -650 M^6 x^2 y^5 z^3 \tau +650 M^6 x\, y^5 z^3 \tau \Big).
\eea
\section{Other molecules spectral functions at LO $\oplus$ $\la\alpha_sG^2\ra$}
\subsection{$\overline{\eta}_q \, \eta_q$ molecule}
\bea
&&\frac{1}{\pi}\mbox{Im}\Pi^{S\,;\, LO}_{\overline{\eta}_q \, \eta_q}(t)= \frac{(12-\epsilon )  }{2048\, \pi ^6} \int_{x\,y\,z} \mathcal{F}_2(M^2,t)\Big[6m^4+4 m^2 y\,z \left(5 t-2 M^2\right)+3 x\, y\, z \left(M^4-6 M^2 t+7 t^2\right) (1-x-y-z)\Big]\nnb \\
&&-\frac{  \epsilon\, m^2 }{256 \,\pi ^6}\int_{x\,y\,z} \mathcal{F}_2(M^2,t)\, y \, z \left(5 t-2 M^2\right), \\
&&\frac{1}{\pi}\mbox{Im}\Pi^{S\,;\,  G^2}_{\overline{\eta}_q \, \eta_q}(t)\hspace*{-0.05cm}=\hspace*{-0.05cm} -\frac{\la g^2 G^2 \ra}{512 \, \pi ^6}  \hspace*{-0.15cm}\int_{x\,y\,z}\hspace*{-0.1cm} \frac{1}{y^3 z} \Big(\hspace*{-0.08cm}4 m^4 x\, z^2\hspace*{-0.1cm}-\hspace*{-0.08cm}6 m^4 y^2\hspace*{-0.1cm}+\hspace*{-0.08cm}4 m^4 y\, z^2\hspace*{-0.1cm}-\hspace*{-0.08cm}6 m^4 y\, z\hspace*{-0.1cm}-\hspace*{-0.08cm}12 m^2 M^2 x^2 y^2\hspace*{-0.1cm}+\hspace*{-0.08cm}12 m^2 M^2 x^2 y\, z^2\hspace*{-0.1cm}-\hspace*{-0.08cm}12 m^2 M^2 x\, y^3\nnb \\
&&+\hspace*{-0.05cm}12 m^2 M^2 x\, y^2 z^2\hspace*{-0.1cm}-\hspace*{-0.08cm}12 m^2 M^2 x\, y^2 z\hspace*{-0.08cm}+\hspace*{-0.08cm}12 m^2 M^2 x\, y^2\hspace*{-0.1cm}+\hspace*{-0.08cm}12 m^2 M^2 x\, y\, z^3\hspace*{-0.1cm}+\hspace*{-0.08cm}6 m^2 M^2 y^3 z\hspace*{-0.08cm}+\hspace*{-0.08cm}18 m^2 t\, x^2 y^2\hspace*{-0.1cm}-\hspace*{-0.08cm}24 m^2 t\, x^2 y\, z^2\hspace*{-0.1cm}+\hspace*{-0.08cm}18 m^2 t\, x\, y^3\nnb \\
&&-24 m^2 t\, x\, y^2 z^2\hspace*{-0.08cm}+\hspace*{-0.08cm}18 m^2 t\, x\, y^2 z\hspace*{-0.08cm}-\hspace*{-0.08cm}18 m^2 t\, x\, y^2\hspace*{-0.08cm}-\hspace*{-0.08cm}24 m^2 t\, x\, y\, z^3\hspace*{-0.08cm}+\hspace*{-0.08cm}6 m^2 t\, x\, y\, z^2\hspace*{-0.08cm}-\hspace*{-0.08cm}9 m^2 t\, y^3 z\hspace*{-0.08cm}+\hspace*{-0.08cm}9 M^4 x^2 y^3 z\hspace*{-0.08cm}+\hspace*{-0.08cm}9 M^4 x\, y^4 z+9 M^4 x\, y^3 z^2\nnb \\
&&-9 M^4 x\, y^3 z\hspace*{-0.08cm}-\hspace*{-0.08cm}36 M^2 t\, x^2 y^3 z\hspace*{-0.08cm}-\hspace*{-0.08cm}36 M^2 t\, x\, y^4 z\hspace*{-0.08cm}-\hspace*{-0.08cm}36 M^2 t\, x\, y^3 z^2\hspace*{-0.08cm}+\hspace*{-0.08cm}36 M^2 t\, x\, y^3 z\hspace*{-0.08cm}+\hspace*{-0.08cm}30 t^2 x^2 y^3 z\hspace*{-0.08cm}+\hspace*{-0.08cm}30 t^2 x\, y^4 z\hspace*{-0.08cm}+\hspace*{-0.08cm}30 t^2 x\, y^3 z^2\hspace*{-0.08cm}-\hspace*{-0.08cm}30 t^2 x\, y^3 z\hspace*{-0.08cm}\Big)\nnb \\
&&+\frac{ \epsilon \, \la g^2 G^2 \ra }{2048 \, \pi ^6} \hspace*{-0.15cm} \int_{x\,y\,z}\hspace*{-0.1cm} \frac{1}{y^3 z}\Big(4 m^4 x\, z^2\hspace*{-0.08cm}-\hspace*{-0.08cm}6 m^4 y^2\hspace*{-0.08cm}+\hspace*{-0.08cm}4 m^4 y\, z^2\hspace*{-0.08cm}-\hspace*{-0.08cm}2 m^4 y\, z\hspace*{-0.08cm}-\hspace*{-0.08cm}8 m^2 M^2 x^2 y^2\hspace*{-0.08cm}+\hspace*{-0.08cm}4 m^2 M^2 x^2 y\, z^2\hspace*{-0.08cm}+\hspace*{-0.08cm}16 m^2 M^2 x\, y^3\hspace*{-0.08cm}+\hspace*{-0.08cm}4 m^2 M^2 x\, y^2 z^2\nnb \\
&&-8 m^2 M^2 x\, y^2 z+8 m^2 M^2 x\, y^2\hspace*{-0.08cm}+\hspace*{-0.08cm}4 m^2 M^2 x\, y\, z^3\hspace*{-0.08cm}+\hspace*{-0.08cm}8 m^2 M^2 x\, y\, z^2+6 m^2 M^2 y^3 z+12 m^2 t\, x^2 y^2-8 m^2 t\, x^2 y\, z^2-24 m^2 t\, x\, y^3\nnb \\
&&-8 m^2 t\, x\, y^2 z^2\hspace*{-0.08cm}+\hspace*{-0.08cm}12 m^2 t\, x\, y^2 z\hspace*{-0.08cm}-\hspace*{-0.08cm}12 m^2 t\, x\, y^2\hspace*{-0.08cm}-\hspace*{-0.08cm}8 m^2 t\, x\, y\, z^3\hspace*{-0.08cm}-\hspace*{-0.08cm}10 m^2 t\, x\, y\, z^2\hspace*{-0.08cm}-\hspace*{-0.08cm}9 m^2 t\, y^3 z\hspace*{-0.08cm}+\hspace*{-0.08cm}9 M^4 x^2 y^3 z+9 M^4 x\, y^4 z+9 M^4 x\, y^3 z^2\nnb \\
&&-\hspace*{-0.05cm}9 M^4 x\, y^3 z\hspace*{-0.08cm}-\hspace*{-0.08cm}36 M^2 t\, x^2 y^3 z\hspace*{-0.08cm}-\hspace*{-0.08cm}36 M^2 t\, x\, y^4 z\hspace*{-0.08cm}-\hspace*{-0.08cm}36 M^2 t\, x\, y^3 z^2\hspace*{-0.08cm}+\hspace*{-0.08cm}36 M^2 t\, x\, y^3 z\hspace*{-0.08cm}+\hspace*{-0.08cm}30 t^2 x^2 y^3 z\hspace*{-0.08cm}+\hspace*{-0.08cm}30 t^2 x\, y^4 z\hspace*{-0.08cm}+\hspace*{-0.08cm}30 t^2 x\, y^3 z^2\hspace*{-0.1cm}-\hspace*{-0.08cm}30 t^2 x\, y^3 z\hspace*{-0.08cm}\Big) \nnb \\
&&-\la g^2 G^2 \ra\Big[\frac{ (12\hspace*{-0.08cm}-\hspace*{-0.08cm}\epsilon )\, m^2 }{3072\, \pi ^6 }\hspace*{-0.15cm} \int_{x\,y\,z} \hspace*{-0.1cm}\frac{1}{y^3} \Big[m^4\hspace*{-0.08cm}+\hspace*{-0.08cm}m^2 t\, z\, (x\hspace*{-0.08cm}+\hspace*{-0.08cm}y)\hspace*{-0.08cm}+\hspace*{-0.08cm}t^2 x\, y\, z\, (1\hspace*{-0.08cm}-\hspace*{-0.08cm}x\hspace*{-0.08cm}-\hspace*{-0.08cm}y\hspace*{-0.08cm}-\hspace*{-0.08cm}z)\Big] \hspace*{-0.08cm}-\hspace*{-0.08cm}\frac{\epsilon \, m^4 }{1536\, \pi ^6 } \hspace*{-0.15cm} \int_{x\,y\,z} \hspace*{-0.1cm}\frac{ t }{y^3}\, z\, \left(x\hspace*{-0.08cm}+\hspace*{-0.08cm}y\right)\Big]\delta(t\hspace*{-0.08cm}-\hspace*{-0.08cm}M^2).
\eea
\subsection{$ \overline{J/\psi}J/\psi,\overline\Upsilon\Upsilon$ molecule}
\bea
&&\frac{1}{\pi}\mbox{Im}\Pi^{S\,;\,LO}_{ \overline{J/\psi}J/\psi,\overline\Upsilon\Upsilon}(t)=\frac{(6+\epsilon) }{256 \,\pi ^6} \hspace*{-0.1cm}\int_{x\,y\,z} \hspace*{-0.15cm} \mathcal{F}_2(M^2,t)  \Big[6 \,m^4\hspace*{-0.08cm}-\hspace*{-0.08cm}2 m^2 y\, z \left(2 M^2-5 t\right)\hspace*{-0.08cm}+\hspace*{-0.08cm}3 x\, y\, z \left(M^4-6 M^2 t+7\, t^2\right)  \left(1-x-y-z\right)\Big], \nnb\\
\\
&&\frac{1}{\pi}\mbox{Im}\Pi^{S\,;\, G^2}_{ \overline{J/\psi}J/\psi,\overline\Upsilon\Upsilon}(t)=\hspace*{-0.1cm}-\frac{ m^2 \la g^2 G^2 \ra}{256 \, \pi ^6 } \hspace*{-0.15cm}\int_{x\,y\,z}\hspace*{-0.1cm} \frac{1}{y^3} \Big(4 m^2 x\, z\hspace*{-0.08cm}+\hspace*{-0.08cm}4 m^2 y\, z\hspace*{-0.08cm}-\hspace*{-0.08cm}12 m^2 y\hspace*{-0.08cm}+\hspace*{-0.08cm}24 M^2 x^2 y\, z\hspace*{-0.08cm}+\hspace*{-0.08cm}24 M^2 x\, y^2 z\hspace*{-0.08cm}+\hspace*{-0.08cm}24 M^2 x\, y\, z^2\hspace*{-0.12cm}\nnb\\
&&-12 M^2 x\, y\, z -6 M^2 y^3-48 t\, x^2 y\, z-48 t\, y^2 z-48 t\, x\, y\, z^2+30 t\, x\, y\, z+9 t\, y^3\Big) \nnb \\
&&-\frac{ \epsilon \, \la g^2 G^2 \ra }{1536 \,\pi ^6 } \hspace*{-0.15cm}\int_{x\,y\,z} \frac{1}{y^3 z} \Big(4 m^4 x\, z^2+12 m^4 y^2+4 m^4 y\, z^2\hspace*{-0.08cm}-\hspace*{-0.08cm}12 m^4 y\, z+24 m^2 M^2 x^2 y\, z^2\hspace*{-0.08cm}-24 m^2 M^2 x\, y^3\hspace*{-0.08cm}+24 m^2 M^2 x\, y^2 z^2\nnb\\
&&+24 m^2 M^2 x\, y\, z^3 -12 m^2 M^2 x\, y\, z^2-18 m^2 M^2 y^3 z-48 m^2 t\, x^2 y\, z^2+36 m^2 t\, x\, y^3-48 m^2 t\, x\, y^2 z^2-48 m^2 t\, x\, y\, z^3\nnb\\
&&+30 m^2 t\, x\, y\, z^2\hspace*{-0.08cm}+\hspace*{-0.08cm}27 m^2 t\, y^3 z\hspace*{-0.08cm}-18 M^4 x^2 y^3 z-18 M^4 x\, y^4 z-18 M^4 x\, y^3 z^2\hspace*{-0.08cm}+\hspace*{-0.08cm}18 M^4 x\, y^3 z\hspace*{-0.08cm}+\hspace*{-0.08cm}72 M^2 t\, x^2 y^3 z\hspace*{-0.08cm}+\hspace*{-0.08cm}72 M^2 t\, x\, y^4 z\nnb \\
&&+72 M^2 t\, x\, y^3 z^2-72 M^2 t\, x\, y^3 z-60 t^2 x^2 y^3 z-60 t^2 x\, y^4 z-60 t^2 x\, y^3 z^2+60 t^2 x\, y^3 z\Big) \nnb \\
&& -\frac{ m^2 \la g^2 G^2 \ra \, (6+\epsilon)}{768\, \pi ^6} \int_{x\,y\,z} \frac{1}{y^3}\Big[2 m^4 + m^2 t\, z\, \left(x + y\right) + 2 t^2 x\, y\, z\, \left(1 - x - y - z\right)\Big] \delta(t-M^2).
\eea
\subsection{ $\overline{\chi} _{1q}\chi _{1q}$ molecule}
\bea
&&\frac{1}{\pi}\mbox{Im}\Pi^{S\,;\, LO}_{\overline{\chi} _{1q}\chi _{1q}}(t)= \frac{3 (6+\epsilon)}{256 \,\pi ^6}\hspace*{-0.15cm}\int_{x\,y\,z}\hspace*{-0.15cm}\mathcal{F}_2(M^2,t) \Big[2 m^4-2 m^2 y\, z \left(5 t-2 M^2\right)+x\, y\, z \left(M^4-6 M^2 t+7 t^2\right) (-x-y-z+1)\Big]\nnb\\
&&+\frac{3 \, m^2}{32 \,\pi ^6}\hspace*{-0.15cm} \int_{x\,y\,z}\hspace*{-0.15cm}\mathcal{F}_2(M^2,t)\, y\,z\left(5 t-2 M^2\right),\\
&&\frac{1}{\pi}\mbox{Im}\Pi^{S\,;\,  G^2}_{\overline{\chi} _{1q}\chi _{1q}}(t)= \frac{  m^2 \la g^2 G^2 \ra }{256\, \pi ^6 } \hspace*{-0.15cm}\int_{x\,y\,z} \hspace*{-0.1cm}\frac{1}{y^3} \Big(4 m^2 x\, z\hspace*{-0.08cm}+\hspace*{-0.08cm}4 m^2 y\, z\hspace*{-0.08cm}+\hspace*{-0.08cm}12 m^2 y\hspace*{-0.08cm}-\hspace*{-0.08cm}24 M^2 x^2 y\, z\hspace*{-0.08cm}-\hspace*{-0.08cm}24 M^2 x\, y^2 z\hspace*{-0.08cm}-\hspace*{-0.08cm}24 M^2 x\, y\, z^2+36 M^2 x\, y\, z\nnb \\
&&-6 M^2 y^3+48 t\, x^2 y\, z+48 t\, x\, y^2 z  +48 t\, x\, y\, z^2-66 t\, x\, y\, z+9 t\, y^3\Big) \nnb \\
&&+\frac{ \epsilon\, \la g^2 G^2 \ra }{512 \, \pi ^6 } \hspace*{-0.15cm}\int_{x\,y\,z}\hspace*{-0.1cm} \frac{1}{y^3 z}\hspace*{-0.08cm} \Big(4 m^4 x\, z^2\hspace*{-0.1cm}-\hspace*{-0.08cm}4 m^4 y^2\hspace*{-0.1cm}+\hspace*{-0.08cm}4 m^4 y\, z^2\hspace*{-0.1cm}+\hspace*{-0.08cm}4 m^4 y\, z\hspace*{-0.08cm}-\hspace*{-0.08cm}8 m^2 M^2 x^2 y\, z^2\hspace*{-0.1cm}+\hspace*{-0.08cm}8 m^2 M^2 x\, y^3\hspace*{-0.1cm}-\hspace*{-0.08cm}8 m^2 M^2 x\, y^2 z^2\hspace*{-0.1cm}-\hspace*{-0.08cm}8 m^2 M^2 x\, y\, z^3 \nnb \\
&&+20 m^2 M^2 x\, y\, z^2-2 m^2 M^2 y^3 z\hspace*{-0.08cm}+\hspace*{-0.08cm}16 m^2 t\, x^2 y\, z^2\hspace*{-0.08cm}-\hspace*{-0.08cm}12 m^2 t\, x\, y^3\hspace*{-0.08cm}+\hspace*{-0.08cm}16 m^2 t\, x\, y^2 z^2\hspace*{-0.08cm}+\hspace*{-0.08cm}16 m^2 t\, x\, y\, z^3-34 m^2 t\, x\, y\, z^2+3 m^2 t\, y^3 z\nnb \\
&&+6 M^4 x^2 y^3 z+6 M^4 x\, y^4 z+6 M^4 x\, y^3 z^2-6 M^4 x\, y^3 z-24 M^2 t\, x^2 y^3 z-24 M^2 t\, x\, y^4 z-24 M^2 t\, x\, y^3 z^2+24 M^2 t\, x\, y^3 z \nnb \\
&&+20 t^2 x^2 y^3 z+20 t^2 x\, y^4 z+20 t^2 x\, y^3 z^2-20 t^2 x\, y^3 z \Big) 
+\hspace*{-0.08cm} \frac{ \epsilon \la g^2 G^2 \ra m^4 }{384 \, \pi ^6 }\hspace*{-0.15cm}  \int_{x\,y\,z} \hspace*{-0.1cm} \frac{ t }{y^3}\,z\, \big(x\hspace*{-0.08cm}+\hspace*{-0.08cm}y\big)  \delta(t\hspace*{-0.08cm}-\hspace*{-0.08cm} M^2)
\nnb \\
 &&-\frac{ m^2\la g^2 G^2 \ra (6\hspace*{-0.08cm}+\hspace*{-0.08cm}\epsilon) }{768 \, \pi ^6 }\hspace*{-0.15cm} \int_{x\,y\,z}\hspace*{-0.1cm} \frac{1}{y^3} \Big[\hspace*{-0.05cm}2 m^4\hspace*{-0.1cm}-\hspace*{-0.1cm}m^2 t\, z \left(x\hspace*{-0.1cm}+\hspace*{-0.1cm}y\right)\hspace*{-0.1cm}+\hspace*{-0.08cm}2 t^2 x\, y\, z \left(1\hspace*{-0.1cm}-\hspace*{-0.1cm}x\hspace*{-0.1cm}-\hspace*{-0.1cm}y\hspace*{-0.1cm}-\hspace*{-0.1cm}z\right)\hspace*{-0.1cm}\Big] \delta(t\hspace*{-0.08cm}-\hspace*{-0.08cm}M^2)\hspace*{-0.08cm}.
\eea
\section{Tetraquarks spectral functions at LO $\oplus$ $\la\alpha_sG^2\ra$}
  
  \subsection{ $\overline S_qS_q$ tetraquark}
\bea
&&\frac{1}{\pi}\mbox{Im}\Pi^{S\,;\, LO}_{\overline S_qS_q}(t)=\frac{1 }{128 \, \pi ^6}  \hspace*{-0.1cm} \int_{x\,y\,z} \hspace*{-0.2cm}\mathcal{F}_2(M^2,t) \; \Big[6 m^4 \hspace*{-0.1cm} + \hspace*{-0.05cm} 4 m^2 \,y \, z \left(5 t-2 M^2\right)+3 x \,y \,z \left(M^4 \hspace*{-0.1cm}-6 M^2 t+7\, t^2\right) \left( 1-x-y-z \right) \hspace*{-0.1cm}\Big], \nnb\\
\\
&&\frac{1}{\pi}\mbox{Im}\Pi^{S\,;\,  G^2}_{\overline S_qS_q}(t)=-\frac{\la g^2 \, G^2 \ra}{1536 \, \pi ^6 } \hspace*{-0.15cm}\int_{x\,y\,z}\hspace*{-0.1cm} \frac{1}{y^3 z} \, \Big\{ 2 m^4 \Big[8 z^2 (x \hspace*{-0.05cm}+ \hspace*{-0.05cm} y)+3 y \,(y-4 \, z)\Big]\hspace*{-0.08cm}+\hspace*{-0.08cm}m^2 y \, \Big[\hspace*{-0.08cm}\left(3 t-2 M^2\right) \hspace*{-0.08cm}\Big(8  x \, z^2 (1 \hspace*{-0.08cm}-\hspace*{-0.08cm}4  x \hspace*{-0.08cm}-\hspace*{-0.08cm}4  y \hspace*{-0.08cm}-4 \, z)\nnb \\
&&+ 6 x \, y \, (1-x-y-z)+3 \, y^2 z\Big) +16  M^2 x \, z^2  \hspace*{0.1cm}(1-x-y-z)\Big]+3  x \, y^3 z \left(3 M^4-12  M^2 t+10 \, t^2\right) \Big(1-x-y-z\Big) \hspace*{-0.1cm}\Big\} \nnb \\
&& -\frac{\la g^2 \, G^2 \ra \, m^2 }{192 \, \pi ^6 }  \int_{x\,y\,z} \frac{1}{y^3} \Big[m^4+m^2 t \,z \, (x+y)+t^2 x \, y \, z \, (1-x-y-z)\Big] \, \delta(t-M^2).
\eea

\subsection{$\overline V_qV_q$ tetraquark}
\bea
&&\frac{1}{\pi}\mbox{Im}\Pi^{S\,;\, LO}_{\overline V_qV_q}(t)=\frac{1 }{32 \, \pi ^6}  \hspace*{-0.1cm} \int_{x\,y\,z}  \hspace*{-0.1cm}\mathcal{F}_2(M^2,t) \, \Big[ 6 \, m^4-2 \, m^2 y \, z \, \left(5 \, t-2 \, M^2\right)+3 \, x \, y \, z\, \hspace*{-0.1cm} \left(M^4-6 \, M^2 t+7 \, t^2\right) (1 \hspace*{-0.1cm}-x \hspace*{-0.1cm}-y \hspace*{-0.1cm}-z)\Big], \nnb\\
\\
&&\frac{1}{\pi}\mbox{Im}\Pi^{S\,;\,  G^2}_{\overline V_qV_q}(t)=\frac{\la g^2 \, G^2 \ra}{768 \, \pi ^6 } \hspace*{-0.15cm}  \int_{x\,y\,z}  \hspace*{0.1cm}\frac{1}{y^3 z}  \hspace*{0.1cm} \Big\{ 2 m^4 \Big[8 \, z^2 (x\hspace*{-0.05cm}+ \hspace*{-0.05cm} y)+3 y \, (5 \, y+8 \, z)\Big]-m^2 y \, \Big[\left(3 \, t-2 \, M^2\right) \Big(3 y^2 (10 \, x+z) \hspace*{-0.1cm}+ \hspace*{-0.1cm}8 x \, z^2\nnb \\
&& \times  \hspace*{0.1cm} (11-8 \, x-8 \, y-8 \, z) \hspace*{0.1cm}\Big) \hspace*{0.1cm}+ \hspace*{0.1cm}32 \, M^2 x \, z^2 \Big(1-x-y-z\Big)\Big]+15 \, x \, y^3 z \, \left(3 \, M^4-12 \, M^2 t+10 \, t^2\right)  \hspace*{0.1cm}\Big(1-x-y-z\Big)\Big\} \nnb \\
&&-  \hspace*{0.1cm} \frac{  m^2 \, \la g^2 \, G^2 \ra }{96 \, \pi ^6 } \int_{x\,y\,z} \frac{ 1}{y^3} \Big[2 \, m^4-m^2 t \, z \, \Big(x+y\Big)+2 \, t^2 x \, y \, z \Big(1-x-y-z\Big)\Big] \, \delta(t-M^2).
\eea 
\subsection{$\bar A_qA_q$ tetraquark (current in Eq.\,\ref{eq:tetra-steele})}
\bea
&&\frac{1}{\pi}\mbox{Im}\Pi^{S\,;\, LO}_{\bar A_qA_q}(t) \hspace*{-0.1cm}=\frac{1 }{64 \, \pi ^6}   \hspace*{-0.1cm}\int_{x\,y\,z}  \hspace*{-0.1cm} \mathcal{F}_2(M^2,t) \, \Big[6 \, m^4 \hspace*{-0.1cm} + \hspace*{-0.1cm} 2 \, m^2 y \, z \, \left(5 \, t-2 \, M^2\right) \hspace*{-0.1cm}+  \hspace*{-0.1cm} 3 \, x \, y \,z \,\left(M^4 \hspace*{-0.1cm}-6 \, M^2 t+7 \, t^2\right) \Big(1 \hspace*{-0.1cm}-x \hspace*{-0.1cm}-y \hspace*{-0.1cm}-z\Big)\Big], \nnb\\
\\
&&\frac{1}{\pi}\mbox{Im}\Pi^{S\,;\,  G^2}_{\bar A_qA_q}(t)=\frac{\la g^2 \, G^2 \ra}{768 \,\pi ^6}  \hspace*{0.1cm} \int_{x\,y\,z}  \hspace*{0.1cm}\frac{1}{y^3 z}  \hspace*{0.1cm}\, \Big\{ m^4 \Big[3 \, y \, \Big(2 \, y+8 \, z\Big)-8 \, z^2 \Big(x+y\Big)\Big]+m^2 y \, \Big[\left(3 \, t-2 \, M^2\right)  \hspace*{0.1cm}\Big(3 \, y^2 (2 \, x-z)\nnb \\
&&-4 \, x \, z^2 (5-8 \, x-8 \, y-8 \, z)\Big)-16 \, M^2 x \, z^2 \Big(1 \hspace*{-0.1cm}-x-y-z\Big)\Big]+3 \, x \, y^3 z \, \left(3 \, M^4-12 \, M^2 t+10 \, t^2\right) \Big(1 \hspace*{-0.1cm}-x-y-z\Big)\Big\}\nnb \\
&& -\frac{m^2 \la g^2 \, G^2 \ra }{192 \, \pi ^6 } \int_{x\,y\,z} \frac{1}{y^3}  \hspace*{0.1cm} \Big[2 \, m^4+m^2 t \, z \, \Big(x+y\Big)+2 \, t^2 x \, y \, z \, \Big(1-x-y-z\Big) \Big] \, \delta(t-M^2).
\eea
\subsection{$\bar A_qA_q$ tetraquark (current in Eq.\,\ref{eq:tetra-wang})}
\bea
&&\frac{1}{\pi}\mbox{Im}\Pi^{S\,;\, LO}_{\bar A_qA_q}(t) \hspace*{-0.1cm}=\frac{1 }{16 \, \pi ^6}   \hspace*{-0.1cm}\int_{x\,y\,z}  \hspace*{-0.1cm} \mathcal{F}_2(M^2,t) \, \Big[6 \, m^4 \hspace*{-0.1cm} + \hspace*{-0.1cm} 2 \, m^2 y \, z \, \left(5 \, t-2 \, M^2\right) \hspace*{-0.1cm}+  \hspace*{-0.1cm} 3 \, x \, y \,z \,\left(M^4 \hspace*{-0.1cm}-6 \, M^2 t+7 \, t^2\right) \Big(1 \hspace*{-0.1cm}-x \hspace*{-0.1cm}-y \hspace*{-0.1cm}-z\Big)\Big],\nnb \\
\\
&&\frac{1}{\pi}\mbox{Im}\Pi^{S\,;\,  G^2}_{\bar A_qA_q}(t)=\frac{\la g^2 \, G^2 \ra}{192 \,\pi ^6}  \hspace*{0.1cm} \int_{x\,y\,z}  \hspace*{0.1cm}\frac{1}{y^3 z}  \hspace*{0.1cm}\, \Big\{ 2 \, m^4 \Big[3 \, y \, \Big( y+4 \, z\Big)-4 \, z^2 \Big(x+y\Big)\Big]+m^2 y \, \Big[\left(3 \, t-2 \, M^2\right)  \hspace*{0.1cm}\Big(3 \, y^2 (2 \, x-z)\nnb \\
&&-4 \, x \, z^2 (5-8 \, x-8 \, y-8 \, z)\Big)-16 \, M^2 x \, z^2 \Big(1 \hspace*{-0.1cm}-x-y-z\Big)\Big]+3 \, x \, y^3 z \, \left(3 \, M^4-12 \, M^2 t+10 \, t^2\right) \Big(1 \hspace*{-0.1cm}-x-y-z\Big)\Big\}\nnb \\
&& -\frac{m^2 \la g^2 \, G^2 \ra }{48 \, \pi ^6 } \int_{x\,y\,z} \frac{1}{y^3}  \hspace*{0.1cm} \Big[2 \, m^4+m^2 t \, z \, \Big(x+y\Big)+2 \, t^2 x \, y \, z \, \Big(1-x-y-z\Big) \Big] \, \delta(t-M^2).
\eea
\section{$\overline P_qP_q$ tetraquark at LO $\oplus$ $G^3$}
\bea
&&\frac{1}{\pi}\mbox{Im}\Pi^{S\,;\, LO}_{\overline P_qP_q}(t)=\frac{1 }{128 \, \pi ^6} \hspace*{-0.15cm} \int_{x\,y\,z} \hspace*{-0.15cm} \mathcal{F}_2(M^2,t)  \,\Big[6 \,  m^4 \hspace*{-0.05cm} - \hspace*{-0.05cm} 4 \,  m^2 y \, z \left(5 \, t \hspace*{-0.05cm} - \hspace*{-0.05cm}2 \, M^2\right)+3 \, x \, y \, z \, \left(M^4\hspace*{-0.1cm}-6 \, M^2 t+7 \, t^2\right)\left(1\hspace*{-0.05cm}- \hspace*{-0.05cm} x \hspace*{-0.05cm}- \hspace*{-0.05cm} y \hspace*{-0.05cm} - \hspace*{-0.05cm} z  \right)\hspace*{-0.05cm}\Big], \nnb\\
\\
&&\frac{1}{\pi}\mbox{Im}\Pi^{S\,;\,  G^2}_{\overline P_qP_q}(t)=\frac{\la g^2 \, G^2 \ra}{1536 \, \pi ^6 } \hspace*{-0.15cm} \int_{x\,y\,z}\frac{1}{y^3 z} \, \Big\{2 \,  m^4 \Big[8 \, z^2 (x \hspace*{-0.05cm}+ \hspace*{-0.05cm} y)-3 \, y^2 \hspace*{-0.05cm}+ \hspace*{-0.05cm} 12 \, y \, z\Big]\hspace*{-0.05cm} -\hspace*{-0.05cm} m^2 y \, \Big[\hspace*{-0.1cm}\left(3 \, t-2 \, M^2\right) \Big(8 \, x \, z^2 (7\hspace*{-0.05cm}-\hspace*{-0.05cm}4 \, x \hspace*{-0.05cm} - \hspace*{-0.05cm} 4 \, y \hspace*{-0.05cm} - \hspace*{-0.05cm} 4 \, z)\nnb \\
&&- 6 \, x \, y \, (1 \hspace*{-0.05cm} - \hspace*{-0.05cm} x \hspace*{-0.05cm} - \hspace*{-0.05cm} y \hspace*{-0.05cm} -\hspace*{-0.05cm} z)-3 \, y^2 z\Big)+16 \, M^2 x  \hspace*{0.1cm} z^2  \hspace*{0.1cm}(1-x-y-z)\Big]-3\, x \, y^3 z \, \left(3 \, M^4-12 \, M^2  \hspace*{0.1cm}t+10 \, t^2\right) (1-x-y-z) \Big\} \nnb \\
&&-\frac{\la g^2 \, G^2 \ra \, m^2 }{192 \, \pi ^6 } \int_{x\,y\,z} \frac{ 1}{y^3} \,\Big[m^4-m^2 \, t \, z  \hspace*{0.1cm}(x+y)+ t^2 x \, y \, z \, (1-x-y-z)\Big]\delta(t-M^2), 
\eea
\bea
&&\frac{1}{\pi}\mbox{Im}\Pi^{S;\,G^3}_{\overline P_qP_q}(t)= -\frac{\la g^3 \, G^3\ra}{221184 \, \pi ^6 }  \int_{x\,y\,z} \frac{1}{y^3 z^2} \, \Big(1848 \, m^2 x^2 y^3+2592 \, m^2 x^2 y^2 z+1408 \, m^2 x^2 y^2+120 \, m^2 x^2 y z^2\nnb \\
&&-260 \, m^2 x^2 y\, z-6912 \, m^2 x^2 z^3\hspace*{+0.05cm}+\hspace*{+0.05cm}1848 \, m^2 x \, y^4\hspace*{+0.05cm}+\hspace*{+0.05cm}4440 \, m^2 x \, y^3 z-440 \, m^2 x \, y^3+2712  \, m^2 x \, y^2 z^2-1444 \, m^2 x \, y^2 z\nnb \\
&&-1408 \, m^2 x \, y^2  -  6792 \, m^2 x \, y \, z^3  -  380 \, m^2 x \, y \, z^2 + \hspace*{0.05cm} 260 \, m^2 x \, y \, z \hspace*{0.05cm} - \hspace*{0.05cm} 6912 \, m^2 x \, z^4 \hspace*{0.05cm} + \hspace*{0.05cm} 13824 \, m^2 x \, z^3 \hspace*{0.05cm} - \hspace*{0.05cm} 1920 \, m^2 y^3 z^2\nnb \\
&&+320 \, m^2 y^3 z \hspace*{-0.08cm}+ \hspace*{-0.08cm} 320 \, m^2 y^2 z^2 \hspace*{-0.08cm}+ \hspace*{-0.08cm} 8160 \, M^2 x^2 y^3 z^2-780 \, M^2 x^2 y^3 z-780 \, M^2 x^2 y^2 z^2-390 \, M^2 x^2 y^2 z+8160 \, M^2 x \, y^4 z^2\nnb \\
&&-780 \, M^2 x \, y^4 z \hspace*{-0.08cm}+ \hspace*{-0.08cm} 8160 \, M^2 x \, y^3 z^3\hspace*{-0.08cm}-\hspace*{-0.08cm}9720 \, M^2 x \, y^3 z^2\hspace*{-0.05cm}+\hspace*{-0.05cm}390 \, M^2 x \, y^3 z \hspace*{-0.08cm}- \hspace*{-0.08cm} 780 \, M^2 x \, y^2 z^3 \hspace*{-0.08cm}+ \hspace*{-0.08cm}390 \, M^2 x \, y^2 z^2+390 \, M^2 x \, y^2 z\nnb \\
&&-20400 \, t \, x^2 y^3 z^2+1560 \, t \, x^2 y^3 z+1560 \, t \, x^2 y^2 z^2+585 \, t \, x^2 y^2 z-20400 \, t \, x \, y^4 z^2+1560 \, t \, x \, y^4 z-\hspace*{0.05cm}20400 \, t \, x \, y^3 z^3\nnb \\
&&+23520 \, t \, x \, y^3 z^2-975 \, t \, x \, y^3 z+1560 \, t \, x \, y^2 z^3-975 \, t \, x \, y^2 z^2-585 \, t \, x \, y^2 z\Big)\nnb \\
&&+\frac{ \la g^3 \, G^3\ra }{552960 \, \pi ^6} \hspace*{-0.1cm} \int_{x\,y\,z}\frac{ e^{-M^2 \tau }}{y^5 z^3} \, \Big(1728 \,  z^3 \tau ^2 m^8 \hspace*{-0.05cm} - \hspace*{-0.05cm} 1728 \, M^2 \, x \, z^4 \tau ^2 m^6 \hspace*{-0.05cm} - \hspace*{-0.05cm} 1728 \, M^2 \, y \, z^4 \, \tau ^2 m^6 \hspace*{-0.05cm}+\hspace*{-0.05cm} 1440 \, y^4 \tau  \, m^6 \hspace*{-0.05cm} - \hspace*{-0.05cm} 10080 \, y \, z^3 \, \tau \,  m^6\nnb \\
&&-200 \, y^3 z \, \tau  \, m^6\hspace*{-0.05cm}+\hspace*{-0.05cm}1440 \, x \, y^5 m^4 \hspace*{0.05cm} - \hspace*{0.05cm} 3456 \, x \, y \, z^5 m^4 \hspace*{0.05cm}+ \hspace*{0.05cm} 1440 \, x^2 y^4 m^4 \hspace*{0.05cm} - \hspace*{0.05cm} 1440 \, x \, y^4 m^4 - \hspace*{0.05cm} 3456 \, x \, y^2 z^4 m^4 + 2880 \, y^2 z^4 m^4 \hspace*{0.05cm} \nnb \\ 
&&- 3456 \, x^2 y \, z^4 m^4 \hspace*{-0.05cm}+ \hspace*{-0.05cm}13536 \, x \, y \, z^4 m^4 \hspace*{-0.1cm}+ \hspace*{-0.05cm} 8640 \, y^2 z^3 m^4 \hspace*{-0.1cm}- \hspace*{-0.05cm} 1080 \, y^4 z^2 m^4 \hspace*{-0.1cm}-\hspace*{-0.05cm}150 \, x \, y^3 z^2 m^4 \hspace*{-0.1cm}+\hspace*{-0.05cm}400 \, y^3 z^2 m^4 \hspace*{-0.1cm}-1728 \, M^4 x \, y \, z^5 \tau ^2 m^4\nnb \\
&&-1728 \, M^4 x \, y^2 z^4 \tau ^2 m^4-1728 \, M^4 x^2 y \, z^4 \tau ^2 m^4+1728 \, M^4 x \, y \, z^4 \tau ^2 m^4-720 \, y^5 z \, m^4 \hspace*{-0.05cm} +1290 \, x \, y^4 z \, m^4-1760 \, y^4 z \, m^4 \nnb \\
&&-150 \, x^2 y^3 z\, m^4 \hspace*{-0.05cm}+\hspace*{-0.05cm}150 \, x \, y^3 z \, m^4 \hspace*{-0.05cm}+\hspace*{-0.05cm}1440 \,  M^2  x \, y^5 \, \tau \,  m^4\hspace*{-0.05cm}- \hspace*{-0.05cm} 3456 \,  M^2  x \, y \, z^5 \, \tau \,  m^4\hspace*{-0.05cm}+1440 \,  M^2  x^2 y^4 \, \tau \, m^4\hspace*{-0.05cm}-1440 \,  M^2  x \, y^4 \, \tau \,  m^4\hspace*{-0.05cm}\nnb \\
&&+2880 \,  M^2  \, y^2 z^4 \, \tau \,  m^4 -3456 \,  M^2  \, x \, y^2  z^4  \tau \,  m^4\hspace*{-0.05cm}- 3456 \, M^2  \, x^2  y \, z^4 \, \tau \,  m^4\hspace*{0.05cm}+\hspace*{0.05cm}13536 \,  M^2  \, x \, y \, z^4 \tau \,  m^4\hspace*{0.05cm}-\hspace*{0.05cm}1080 \,  M^2  \, y^4 z^2 \tau \, m^4\hspace*{0.05cm}\nnb \\
&& -\hspace*{0.05cm}150 \,  M^2  \, x \, y^3 \hspace*{0.05cm} z^2 \tau \,  m^4 \hspace*{0.05cm}- \hspace*{0.05cm}720 \,  M^2 \, y^5 \hspace*{0.05cm} z \, \tau \,  m^4 \hspace*{0.05cm} + \hspace*{0.05cm} 1290 \,  M^2 \, x \, y^4 \hspace*{0.05cm} z \, \tau \,  m^4 \hspace*{0.05cm} -  \hspace*{0.05cm}150 \,  M^2 \, x^2 y^3 z \, \tau \,  m^4 \hspace*{0.05cm}+ \hspace*{0.05cm} 150 \,  M^2 \, x \, y^3 z \, \tau \,  m^4 \nnb \\
&&+  11520 \,  M^2 \, x \, y^2 z^5 m^2\hspace*{0.05cm}+ \hspace*{0.05cm}11320 \,  M^2 \,  x \, y^3 z^4 m^2 \hspace*{0.05cm}+ \hspace*{0.05cm} 11520 \,  M^2 \, x^2 \hspace*{0.05cm} y^2 z^4 m^2\hspace*{0.05cm} -\hspace*{0.05cm}20160 \,  M^2 \,  x \, y^2\hspace*{0.05cm} z^4 m^2 \hspace*{0.05cm} +\hspace*{0.05cm} 3200 \,  M^2 \, y^5 \hspace*{0.05cm} z^3 m^2 \nnb \\
&& - 400 \,  M^2 \, y^4 z^3 m^2 \hspace*{-0.08cm} -\hspace*{-0.08cm}4520 \,  M^2 \, x \,  y^4 z^3 m^2\hspace*{-0.08cm} -\hspace*{-0.05cm}200 \,  M^2 x^2 y^3 z^3 m^2 \hspace*{-0.08cm} +\hspace*{-0.08cm}525 \,  M^2   x \, y^3 z^3 m^2 \hspace*{-0.08cm} -\hspace*{-0.08cm}400 \,  M^2  y^5 z^2 m^2\hspace*{-0.08cm}-\hspace*{-0.08cm}7400 \,  M^2  x \, y^5 z^2 m^2 \nnb \\ 
&&-4320 \,  M^2 \,  x^2 y^4 \hspace*{0.05cm} z^2 m^2\hspace*{0.05cm}+\hspace*{0.05cm}2885 \,  M^2 \, x \, y^4 \hspace*{0.05cm} z^2 m^2\hspace*{0.05cm}+\hspace*{0.05cm}325 \,  M^2 \, x^2 \hspace*{0.05cm} y^3 \hspace*{0.05cm} z^2 m^2\hspace*{0.05cm}-\hspace*{0.05cm}325 \,  M^2 \, x \, y^3 \hspace*{0.05cm} z^2 \hspace*{0.05cm} m^2\hspace*{0.05cm}-\hspace*{0.05cm}3080 \,  M^2 \, x \, y^6 \hspace*{0.05cm} z \, m^2\hspace*{0.05cm}\nnb \\
&& -\hspace*{0.05cm}3080 \,  M^2 \, x^2 \hspace*{0.05cm} y^5 \hspace*{0.05cm} z \, m^2\hspace*{0.05cm}+\hspace*{0.05cm}1320 \,  M^2  x \, y^5 \hspace*{0.05cm} z \, m^2 \hspace*{0.05cm} - \hspace*{0.05cm} 1760 \,  M^2 \, x^2 y^4 z m^2 \hspace*{0.05cm}+ \hspace*{0.05cm} 1760 \,  M^2 \, x \, y^4 z \, m^2 \hspace*{0.05cm} + \hspace*{0.05cm} 2880 \, M^4 x \, y^2 z^5 \tau \, m^2 \nnb \\
&&+2830 \, M^4 \, x \, y^3 z^4 \tau \,  m^2 \hspace*{0.05cm} + \hspace*{0.05cm} 2880 \, M^4 \, x^2 y^2 z^4 \tau \,  m^2 \hspace*{0.05cm} - \hspace*{0.05cm} 2880 \, M^4 \, x \, y^2 z^4 \tau \, m^2 \hspace*{0.05cm} +\hspace*{0.05cm}800 \, M^4 \, y^5 z^3 \tau \, m^2 - 1130 \, M^4  x \, y^4 z^3 \tau \,  m^2 \nnb \\
&&-50 \, M^4 x^2 \hspace*{0.05cm} y^3 \hspace*{0.05cm} z^3 \tau \,  m^2 \hspace*{0.05cm}+ \hspace*{0.05cm} 50 \, M^4 x \, y^3 \hspace*{0.05cm} z^3 \tau \,  m^2  \hspace*{0.05cm}- \hspace*{0.05cm} 1850 \, M^4 x \, y^5 \hspace*{0.05cm} z^2 \hspace*{0.05cm} \tau \,  m^2\hspace*{0.05cm} - \hspace*{0.05cm} 1080 \, M^4 x^2 \hspace*{0.05cm} y^4 \hspace*{0.05cm} z^2 \tau \,  m^2\hspace*{0.05cm}+ \hspace*{0.05cm} 1080 \, M^4 x \, y^4 z^2 \tau  \, m^2\nnb \\
&&-770 \, M^4 x \, y^6 z \, \tau \,  m^2 -770 \, M^4 x^2 y^5 z \, \tau \, m^2 \hspace*{-0.05cm}+\hspace*{-0.05cm} 770 \, M^4 x \, y^5 z \, \tau  \, m^2+ 7650 \, M^4 x \, y^5 z^4  -  325 \, M^4 x \, y^4 z^4   +  7650 \, M^4 x \, y^6 z^3\nnb \\
&& + 7650 \, M^4 x^2 y^5 z^3 \hspace*{-0.1cm}  - \hspace*{-0.05cm} 8300 \, M^4 x \, y^5 z^3 \hspace*{-0.1cm} -\hspace*{-0.05cm} 325 \, M^4 x^2 y^4 z^3 \hspace*{-0.1cm}+\hspace*{-0.05cm}325 \, M^4 x \, y^4 z^3\hspace*{-0.1cm} -\hspace*{-0.05cm} 325 \, M^4 x \, y^6 z^2\hspace*{-0.1cm} - \hspace*{-0.05cm}325 \, M^4 x^2 y^5 z^2 \hspace*{-0.1cm} + \hspace*{-0.05cm} 325 \, M^4 x \, y^5 z^2 \nnb \\
&&+850 \, M^6 x \, y^5 z^4 +850 \, M^6 x \, y^6 z^3 \tau +850 \, M^6 x^2 y^5 z^3 \tau  -850 \, M^6 x \, y^5 z^3 \tau \Big).
\eea
 \end{widetext}
\newpage
 \section*{References}
\input{bib_4Q.tex}
 \end{document}
